\documentclass[12pt,preprint]{aastex62}

\usepackage{amssymb}
\usepackage{amsmath}
\usepackage{epstopdf}
\usepackage{color}
\usepackage{amsmath}
\usepackage{graphicx}
\usepackage{hyperref}
\usepackage{graphics,graphicx,natbib}
\usepackage{cleveref}

\def\lapp{\ifmmode\stackrel{<}{_{\sim}}\else$\stackrel{<}{_{\sim}}$\fi}
\def\gapp{\ifmmode\stackrel{>}{_{\sim}}\else$\stackrel{>}{_{\sim}}$\fi}

\newcommand{\degrees}{\ensuremath{^{\circ}}}

\newcommand{\st}{}%{\color{green}}
\newcommand{\ef}{}

\begin{document}
\title{Nine New Repeating Fast Radio Burst Sources from CHIME/FRB} 
\shorttitle{Nine New Repeaters from CHIME/FRB}
\shortauthors{}

\author{E.~Fonseca}
    \affiliation{Department of Physics, McGill University, 3600 rue University, Montr\'eal, QC H3A 2T8, Canada}
    \affiliation{McGill Space Institute, McGill University, 3550 rue University, Montr\'eal, QC H3A 2A7, Canada}
\author{B.~C.~Andersen}
    \affiliation{Department of Physics, McGill University, 3600 rue University, Montr\'eal, QC H3A 2T8, Canada}
    \affiliation{McGill Space Institute, McGill University, 3550 rue University, Montr\'eal, QC H3A 2A7, Canada}
\author{M.~Bhardwaj}
    \affiliation{Department of Physics, McGill University, 3600 rue University, Montr\'eal, QC H3A 2T8, Canada}
    \affiliation{McGill Space Institute, McGill University, 3550 rue University, Montr\'eal, QC H3A 2A7, Canada}
\author{P.~Chawla}
    \affiliation{Department of Physics, McGill University, 3600 rue University, Montr\'eal, QC H3A 2T8, Canada}
    \affiliation{McGill Space Institute, McGill University, 3550 rue University, Montr\'eal, QC H3A 2A7, Canada}
\author{D.~C.~Good}
    \affiliation{Department of Physics and Astronomy, University of British Columbia, 6224 Agricultural Road, Vancouver, BC V6T 1Z1, Canada}
\author{A.~Josephy}
    \affiliation{Department of Physics, McGill University, 3600 rue University, Montr\'eal, QC H3A 2T8, Canada}
    \affiliation{McGill Space Institute, McGill University, 3550 rue University, Montr\'eal, QC H3A 2A7, Canada}
\author{V.~M.~Kaspi}
    \affiliation{Department of Physics, McGill University, 3600 rue University, Montr\'eal, QC H3A 2T8, Canada}
    \affiliation{McGill Space Institute, McGill University, 3550 rue University, Montr\'eal, QC H3A 2A7, Canada}
\author{K.~W.~Masui}
    \affiliation{MIT Kavli Institute for Astrophysics and Space Research, Massachusetts Institute of Technology, 77 Massachusetts Ave, Cambridge, MA 02139, USA}
    \affiliation{Department of Physics, Massachusetts Institute of Technology, 77 Massachusetts Ave, Cambridge, MA 02139, USA}
\author{R.~Mckinven}
    \affiliation{Dunlap Institute for Astronomy and Astrophysics, University of Toronto, 50 St. George Street, Toronto, ON M5S 3H4, Canada}
    \affiliation{David A. Dunlap Department of Astronomy and Astrophysics, University of Toronto, 50 St. George Street, Toronto, ON M5S 3H4, Canada}
\author{D.~Michilli}
    \affiliation{Department of Physics, McGill University, 3600 rue University, Montr\'eal, QC H3A 2T8, Canada}
    \affiliation{McGill Space Institute, McGill University, 3550 rue University, Montr\'eal, QC H3A 2A7, Canada}
\author{Z.~Pleunis}
    \affiliation{Department of Physics, McGill University, 3600 rue University, Montr\'eal, QC H3A 2T8, Canada}
    \affiliation{McGill Space Institute, McGill University, 3550 rue University, Montr\'eal, QC H3A 2A7, Canada}
\author{K.~Shin}
    \affiliation{MIT Kavli Institute for Astrophysics and Space Research, Massachusetts Institute of Technology, 77 Massachusetts Ave, Cambridge, MA 02139, USA}
    \affiliation{Department of Physics, Massachusetts Institute of Technology, 77 Massachusetts Ave, Cambridge, MA 02139, USA}
\author{S.~P.~Tendulkar}
    \affiliation{Department of Physics, McGill University, 3600 rue University, Montr\'eal, QC H3A 2T8, Canada}
\author{K.~M.~Bandura}
    \affiliation{CSEE, West Virginia University, Morgantown, WV 26505, USA}
    \affiliation{Center for Gravitational Waves and Cosmology, West Virginia University, Morgantown, WV 26505, USA}
\author{P.~J.~Boyle}
    \affiliation{Department of Physics, McGill University, 3600 rue University, Montr\'eal, QC H3A 2T8, Canada}
    \affiliation{McGill Space Institute, McGill University, 3550 rue University, Montr\'eal, QC H3A 2A7, Canada}
\author{C.~Brar}
    \affiliation{Department of Physics, McGill University, 3600 rue University, Montr\'eal, QC H3A 2T8, Canada} 
    \affiliation{McGill Space Institute, McGill University, 3550 rue University, Montr\'eal, QC H3A 2A7, Canada}
\author{T.~Cassanelli}
    \affiliation{Dunlap Institute for Astronomy and Astrophysics, University of Toronto, 50 St. George Street, Toronto, ON M5S 3H4, Canada} 
    \affiliation{David A. Dunlap Department of Astronomy and Astrophysics, University of Toronto, 50 St. George Street, Toronto, ON M5S 3H4, Canada}
\author{D.~Cubranic}
    \affiliation{Department of Physics and Astronomy, University of British Columbia, 6224 Agricultural Road, Vancouver, BC V6T 1Z1, Canada}
\author{M.~Dobbs}
    \affiliation{Department of Physics, McGill University, 3600 rue University, Montr\'eal, QC H3A 2T8, Canada}
    \affiliation{McGill Space Institute, McGill University, 3550 rue University, Montr\'eal, QC H3A 2A7, Canada}
\author{F.~Q.~Dong}
    \affiliation{Department of Physics and Astronomy, University of British Columbia, 6224 Agricultural Road, Vancouver, BC V6T 1Z1, Canada}
\author{B.~M.~Gaensler}
    \affiliation{Dunlap Institute for Astronomy and Astrophysics, University of Toronto, 50 St. George Street, Toronto, ON M5S 3H4, Canada}
    \affiliation{David A. Dunlap Department of Astronomy and Astrophysics, University of Toronto, 50 St. George Street, Toronto, ON M5S 3H4, Canada}
\author{G.~Hinshaw}
    \affiliation{Department of Physics and Astronomy, University of British Columbia, 6224 Agricultural Road, Vancouver, BC V6T 1Z1, Canada}
\author{T.~L.~Landecker}
    \affiliation{Dominion Radio Astrophysical Observatory, Herzberg Astronomy and Astrophysics Research Centre, National Research Council Canada, P.O. Box 248, Penticton, BC V2A 6J9, Canada}
\author{C.~Leung}
    \affiliation{MIT Kavli Institute for Astrophysics and Space Research, Massachusetts Institute of Technology, 77 Massachusetts Ave, Cambridge, MA 02139, USA}
    \affiliation{Department of Physics, Massachusetts Institute of Technology, 77 Massachusetts Ave, Cambridge, MA 02139, USA}
\author{D.~Z.~Li}
    \affiliation{Department of Physics, University of Toronto, 60 St. George Street, Toronto, ON M5S 1A7, Canada} 
    \affiliation{Canadian Institute for Theoretical Astrophysics, University of Toronto, 60 St. George Street, Toronto, ON M5S 3H8, Canada}
    \affiliation{Max Planck Institute for Radio Astronomy, Auf dem Huegel 69, 53121 Bonn, Germany}
\author{H.~-H.~Lin}
    \affiliation{Canadian Institute for Theoretical Astrophysics, University of Toronto, 60 St. George Street, Toronto, ON M5S 3H8, Canada}
    \affiliation{Max Planck Institute for Radio Astronomy, Auf dem Huegel 69, 53121 Bonn, Germany}
\author{J.~Mena-Parra}
    \affiliation{MIT Kavli Institute for Astrophysics and Space Research, Massachusetts Institute of Technology, 77 Massachusetts Ave, Cambridge, MA 02139, USA}
\author{M.~Merryfield}
    \affiliation{Department of Physics, McGill University, 3600 rue University, Montr\'eal, QC H3A 2T8, Canada}
    \affiliation{McGill Space Institute, McGill University, 3550 rue University, Montr\'eal, QC H3A 2A7, Canada}
\author{A.~Naidu}
    \affiliation{Department of Physics, McGill University, 3600 rue University, Montr\'eal, QC H3A 2T8, Canada}
    \affiliation{McGill Space Institute, McGill University, 3550 rue University, Montr\'eal, QC H3A 2A7, Canada}
\author{C.~Ng}
    \affiliation{Dunlap Institute for Astronomy and Astrophysics, University of Toronto, 50 St. George Street, Toronto, ON M5S 3H4, Canada}
\author{C.~Patel}
    \affiliation{Dunlap Institute for Astronomy and Astrophysics, University of Toronto, 50 St. George Street, Toronto, ON M5S 3H4, Canada}
    \affiliation{Department of Physics, McGill University, 3600 rue University, Montr\'eal, QC H3A 2T8, Canada}
\author{U.~Pen}
    \affiliation{Canadian Institute for Theoretical Astrophysics, University of Toronto, 60 St. George Street, Toronto, ON M5S 3H8, Canada}
    \affiliation{Dunlap Institute for Astronomy and Astrophysics, University of Toronto, 50 St. George Street, Toronto, ON M5S 3H4, Canada}
    \affiliation{Canadian Institute for Advanced Research, CIFAR Program in Gravitation and Cosmology, Toronto, ON M5G 1Z8, Canada}
    \affiliation{Perimeter Institute for Theoretical Physics, 31 Caroline Street North, Waterloo, ON N2L 2Y5, Canada}
    \affiliation{Max Planck Institute for Radio Astronomy, Auf dem Huegel 69, 53121 Bonn, Germany}
\author{M.~Rafiei-Ravandi}
    \affiliation{Perimeter Institute for Theoretical Physics, 31 Caroline Street North, Waterloo, ON N2L 2Y5, Canada}
\author{M.~Rahman}
    \affiliation{Dunlap Institute for Astronomy and Astrophysics, University of Toronto, 50 St. George Street, Toronto, ON M5S 3H4, Canada}
\author{S.~M.~Ransom}
    \affiliation{National Radio Astronomy Observatory, 520 Edgemont Rd., Charlottesville, VA 22903, USA}
\author{P.~Scholz}
    \affiliation{Dunlap Institute for Astronomy and Astrophysics, University of Toronto, 50 St. George Street, Toronto, ON M5S 3H4, Canada}
    \affiliation{Dominion Radio Astrophysical Observatory, Herzberg Astronomy and Astrophysics Research Centre, National Research Council Canada, P.O. Box 248, Penticton, BC V2A 6J9, Canada}
\author{K.~M.~Smith}
    \affiliation{Perimeter Institute for Theoretical Physics, 31 Caroline Street North, Waterloo, ON N2L 2Y5, Canada}
\author{I.~H.~Stairs}
    \affiliation{Department of Physics and Astronomy, University of British Columbia, 6224 Agricultural Road, Vancouver, BC V6T 1Z1, Canada}
\author{K.~Vanderlinde}
    \affiliation{Dunlap Institute for Astronomy and Astrophysics, University of Toronto, 50 St. George Street, Toronto, ON M5S 3H4, Canada} 
    \affiliation{David A. Dunlap Department of Astronomy and Astrophysics, University of Toronto, 50 St. George Street, Toronto, ON M5S 3H4, Canada}
\author{P.~Yadav}
    \affiliation{Department of Physics and Astronomy, University of British Columbia, 6224 Agricultural Road, Vancouver, BC V6T 1Z1, Canada}
\author{A.~V.~Zwaniga}
    \affiliation{Department of Physics, McGill University, 3600 rue University, Montr\'eal, QC H3A 2T8, Canada}
    \affiliation{McGill Space Institute, McGill University, 3550 rue University, Montr\'eal, QC H3A 2A7, Canada}

\correspondingauthor{E. Fonseca}
\email{efonseca@physics.mcgill.ca}

\begin{abstract}
    We report on the discovery and analysis of bursts from nine new repeating fast radio burst (FRB) sources found using the Canadian Hydrogen Intensity Mapping Experiment (CHIME) telescope. These sources span a dispersion measure (DM) range of 195 to 1380 pc cm$^{-3}$. We detect two bursts from three of the new sources, three bursts from four of the new sources, four bursts from one new source, and five bursts from one new source. We determine sky coordinates of all sources with uncertainties of $\sim$10$^\prime$. We detect Faraday rotation measures for two sources, with values $-20(1)$ and $-499.8(7)$~rad~m$^{-2}$, that are substantially lower than the RM derived from bursts emitted by FRB 121102. We find that the DM distribution of our events, combined with the nine other repeaters discovered by CHIME/FRB, is indistinguishable from that of thus far non-repeating CHIME/FRB events. However, as previously reported, the burst widths appear statistically significantly larger than the thus far non-repeating CHIME/FRB events, further supporting the notion of inherently different emission mechanisms and/or local environments. These results are consistent with previous work, though are now derived from 18 repeating sources discovered by CHIME/FRB during its first year of operation. We identify candidate galaxies that may contain FRB 190303.J1353+48 (DM = 222.4 pc cm$^{-3}$).
\end{abstract}
 
%\keywords{pulsars: general --- pulsars: timing --- surveys, RRATs}

\section{Introduction}
FRBs are an enigmatic class of radio transients that exhibit millisecond durations, cosmological distances and large energy output \citep[$\sim10^{40}$ erg;][]{dgbb15}. Such extreme characteristics have resulted in a diverse and evolving landscape of physical models that differ in progenitor and emission-mechanism types, with possible interpretations ranging from compact objects to cosmic strings \citep[for a living summary of proposed models\footnote{https://frbtheorycat.org}, see][]{pww+18}.

Among the most intriguing FRB sources are those that emit multiple bursts, as first seen in FRB 121102 \citep{ssh+16a,ssh+16b}. Repeating sources necessitate non-cataclysmic models. Recently proposed models typically include young, highly magnetised neutron stars that likely interact with their environments, such as supernova remnants \citep[e.g.][]{lyu14,mkm16,bel17,mms19} or massive black holes \citep[e.g.][]{zhang2018frb}, or bursting due to internal magnetar instabilities affecting the magnetosphere \citep[e.g.][]{lyu19}. No model yet fully explains all existing, albeit limited, observational data. Therefore, discoveries of new repeating FRBs are important to constrain available models.

Interferometric follow-up observations of bursts can be performed to obtain precise sub-arcsecond positions, which can then be observed with optical telescopes to identify host galaxies and their redshifts. The localization of FRB 121102 through direct imaging of repeat bursts \citep{clw+17, mph+17} pinpointed the source to a star-forming region in a dwarf galaxy at redshift $z = 0.193$ \citep{tbc+17,bassa2017frb}, enabling  multi-wavelength studies of the source environment \citep[e.g.][]{sbh+17}. Recent localizations of single-burst FRBs show that FRBs can also reside in more massive galaxies with a wide range of specific star formation rates \citep{bdp+19,rcd+19,pmm+19,mnh+20}. Additional localizations and subsequent multi-wavelength observations will provide a unique data set of host classifications, redshifts, assessments of source environments and other information.

The CHIME telescope and its FRB search backend \citep{abb+18} first detected 13 low-frequency bursts \citep{abb+19a}, and discovered the second repeating FRB source from this initial sample \citep{abb+19b}. CHIME/FRB recently published a detection of FRB 121102 in the CHIME band \citep{jcf+19}, as well as eight new repeating sources whose bursts generally show complex morphological features and, as an ensemble, possess larger widths than those of thus far non-repeating FRBs \citep[][hereafter Paper I]{abb+19c}. The latter observation serves as significant evidence of different emission mechanisms between repeating sources and apparent single-burst FRB sources, and/or common environments surrounding repeating sources, though a larger sample of both FRB types will be helpful for verifying this apparent trend. Moreover, the observed Faraday rotation measure (RM) for the CHIME/FRB repeater FRB 180916.J0158+65 ($-114.6 \pm 0.6$ rad m$^{-2}$; Paper I) is considerably smaller than the RM$\sim 10^5$ rad m$^{-2}$ measured for FRB 121102 \citep{msh+18}; moreover, FRB J180916.J0518+65 has no persistent radio counterpart \citep{mnh+20} while FRB 121102 is coincident with a persistent source \citep{mph+17}. These facts suggest that dramatically different magneto-ionic environments can contain FRB sources.

In this work, we report the discovery and subsequent analyses of bursts emitted from nine new repeating FRB sources. In Section \ref{sec:obs}, we highlight the observations taken with the CHIME telescope that enabled offline analyses of total-intensity and polarization spectra. In Section \ref{sec:analysis}, we describe the analysis used for quantifying per-burst and sample properties.  In Section \ref{sec:discussion}, we discuss the results obtained from analyses of burst morphology, sample properties, baseband data and multi-wavelength counterparts. In Section \ref{sec:conclusions}, we summarise our findings.

\section{Observations} 
\label{sec:obs}

All detections presented here were made during a period of telescope commissioning between 28 August 2018 and 30 September 2019 using the CHIME/FRB system. As described by \citet{abb+18}, the CHIME/FRB backend continuously receives a total-intensity, polarization-summed time series generated by a FX correlator with 0.98304 ms cadence for 16,384 frequency channels across the 400--800 MHz band. The 128-node FRB system employs real-time radio frequency interference (RFI) mitigation and a modified tree de-dispersion algorithm. Candidate signals with integrated S/N values greater than a configurable threshold are immediately forwarded to a post-detection pipeline\footnote{Documentation for the post-detection part of the CHIME/FRB realtime pipeline can be found at \url{https://chimefrb.github.io/frb-l2l3/}} for real-time  classification to i) identify and ignore RFI-induced signals, ii) check for coincidence with known Galactic sources and iii) compare with predicted Galactic contributions to DM. Signals are classified as extragalactic (i.e., FRBs) if they are not associated with any known Galactic sources, and their measured DMs exceed independent estimates of the maximum values predicted by available Galactic DM models \citep{ne2001,ymw17}; we do not account for DM contributions from electronic content in the Galactic halo.

We deemed a group of bursts to originate from the same repeating source if their measured position and DM values lied close to one another, such that the differences in these quantities satisfied the following criteria given estimated uncertainties: $\Delta$DM $<$ 1 pc cm$^{-3}$; $\Delta$R.A.$\cos$(Dec.) $<$ 1 deg.; and $\Delta$Dec. $<$ 1 deg. These thresholds were chosen based on a statistical analysis of chance coincidence in the presence of a large FRB sample; see Appendix \ref{sec:chance} for a discussion of simulations that motivate these bounds for CHIME/FRB, given that $\sim$700 FRBs have been detected with the CHIME/FRB system during the aforementioned time period.\footnote{Analysis of the $\sim$700 FRBs found by CHIME/FRB is ongoing and will be published in a forthcoming catalog.} The thresholds used in this work are more stringent than those used in Paper I, with the DM threshold being increased by an order of magnitude.

The CHIME/FRB real-time pipeline automatically records 60-s segments of total-intensity data to disk at the resolution used by the FRB instrument for detection, for all burst events deemed astrophysical and extragalactic. For sufficiently bright signals, the CHIME/FRB system also records buffered telescope baseband data --- complex voltages measured with 4-bit precision (each for real and imaginary parts) by all 1,024 dual-polarization feeds at 2.56-$\mu$s cadence across 1,024 frequency channels --- which allows for offline polarization detection, position verification, and high-time-resolution studies of burst morphology. For this work, we analyzed polarization properties and verified localizations using available baseband data; we deferred morphological studies with baseband data for future study as analysing $\mu$s-level structure is beyond the scope of this work.

Since Paper I, we lowered the S/N thresholds for recording intensity and baseband data to disk as confidence in the real-time detection system grew. For repeating sources, total-intensity spectra of bursts presented here are drawn from a wider S/N distribution, going down to a real-time detection S/N of 8 instead of 9. A larger number of bursts were detected with 9 $<$ S/N $<$ 10 than in Paper I due to improved system sensitivity and classification algorithms during the telescope commissioning period. For baseband recording, we used a threshold S/N of 9.

\begin{table}[t]
\begin{center}
\caption{Properties of Nine New CHIME/FRB Repeating Sources}
%\footnotesize
\centering
\resizebox{1.05\textwidth}{!}{ 
\hspace{-1.8in}
\begin{tabular}{cccccccccccc} \hline
%    Nickname$^a$ & Name$^b$  &  R.A.$^c$ & Dec.$^c$  & No. Bursts & Exposure (hr) & Completeness (Jy ms) \\
    Source & Name$^a$  &  R.A.$^b$ & Dec.$^b$ & $l^c$ & $b^c$ & DM$^d$ & DM$_{\rm NE2001}^e$ & DM$_{\rm YMW16}^e$ & N$_{\rm bursts}$  & Exposure$^f$ & Completeness$\,^g$  \\
          &       &  (J2000)    &  (J2000) & (deg) & (deg) & (pc~cm$^{-3}$) & (pc~cm$^{-3}$) &(pc~cm$^{-3}$) &      &     (hr, upper / lower) &  (Jy ms)  \\\hline
1 &  190208.J1855+46 & {\it 18h55m$\pm$14$'$}   & {\it +46$^\circ$58$'\pm$15$'$} & 76.8 & 18.9 & 580.05(15) & 72 & 66 & 2 & 20$\pm$14 & 3.4 \\
2 &  190604.J1435+53 & 14h35m$\pm$10$'$  & +53$^\circ$17$'\pm$11$'$ & 93.8 & 57.6 & 552.65(5) & 32 & 24 & 2 & 30$\pm$11& 2.8\\
3 &  190212.J18+81 & 18h24m$\pm$15$'$  & +81$^\circ$26$'\pm$10$'$ & 113.3 & 27.8 & 302(1) & 49 & 44 & 3$^h$ & 55$\pm$52 / 159$\pm$11 & 8.2 / 13\\
 & & 17h39m$\pm$16$'$ & +81$^\circ$24$'\pm$7$'$ & 113.5 & 29.5 &  & & & & & \\
4 & 180908.J1232+74 & 12h32m$\pm$17$'$  & +74$^\circ$12$'\pm$19$'$ & 124.7  & 42.9 & 195.6(2)  & 38 & 31 & 4 & 53$\pm$33 / 36$\pm$25 & 5.9 / 18\\
5 &  190117.J2207+17 & 22h07m$\pm$8$'$  & +17$^\circ$23$'\pm$15$'$ & 76.4 & $-$30.3 & 393.6(8) & 48 & 40 & 5 & 19$\pm$8& 6.5\\
6 &  190303.J1353+48 & 13h53m$\pm$14$'$ & +48$^\circ$15$'\pm$15$'$ & 97.5 & 65.7 & 222.4(7) & 29 & 22 & 3 & 23$\pm$12 & 2.6\\
%R10$^a$ & 180730.JXXXX+87  & & &  2 & 157$\pm$71& 184$\pm$66\\
7 & 190417.J1939+59 & {\it 19h39m$\pm$13$'$}   & {\it +59$^\circ$24$'\pm$16$'$} & 91.5 & 17.4 & 1378.2(2) & 78 & 80 & 3 & 29$\pm$19 & 4.3 \\
8 & 190213.J02+20 & 02h14m$\pm$16$'$ & +20$^\circ$04$'\pm$20$'$ & 148.1 & $-$38.7 & 651.45(5) & 43 & 34 & 2 & 17$\pm$9& 4.4\\
 & & 02h07m$\pm$16$'$ & +20$^\circ$05$'\pm$20$'$ & 146.1 & $-$39.4 & & & & & & \\
%9 & 190303.J1221+70 & 12h21m$\pm$15$'$ & +70$^\circ$51$'\pm$17$'$ & 126.5 & 46.1 & 714(2) & 37 & 29 & 2 & 40$\pm$29 / 48$\pm$10 & 1.4 / 4.8 \\
9 &  190907.J08+46 & 08h09m$\pm$11$'$ & +46$^\circ$16$'\pm$14$'$ & 173.4 & 32.3 & 309.6(2) & 53 & 51 & 3 & 23$\pm$14& 2.5\\
 & & 08h02m$\pm$12$'$ & +46$^\circ$15$'\pm$14$'$ & 173.2 & 31.1 & & & & & & \\\hline
\end{tabular}
}
\label{ta:repeaters}
\end{center}

$^a$ Here we employ the naming convention (YYMMDD.JHHMM$\pm$DD) used in \cite{abb+19a} and \cite{abb+19b} in the current absence of a final naming convention agreed upon by the community.  These names therefore are likely to change.  The date in the name corresponds to our first detection of the source.  For brevity, and for the remainder of the paper, we refer to the repeaters by Source number (Column 1). For sources with non-contiguous error regions in Fig ure~\ref{fig:localization}, the name is defined by the central position, except for Sources 3, 8, and 9, for which the `central' R.A. is not well defined at the minute level.\\
$^b$ Positions were determined from per-burst S/N data (see Section \ref{sec:localization}). Sources with position in italics have three or more non-contiguous error regions, with the tabulated position referring to the central region, with 90\% confidence uncertainty regions.  See Figure~\ref{fig:localization} for details. Sources 3, 8, and 9 have two non-contiguous uncertainty regions, resulting in two position entries (see Fig.~\ref{fig:localization}).\\
$^c$ Galactic longitude and latitude for the best position. \\
$^d$ Weighted average DM (see Table~\ref{ta:bursts}).\\
$^e$ Maximum model prediction along this line-of-sight for the NE2001 \citep{ne2001} and YMW16 \citep{ymw17} Galactic electron density distribution models. Neither model accounts for DM contributions from the Galactic halo, which is thought to be 50--80 pc cm$^{-3}$ \citep{prochaska2019probing}.\\
$^f$ For sources observed twice a day, the second entry corresponds to the less sensitive lower transit. The uncertainties in the total exposure for the upper and lower transits of each source are dominated by the corresponding source declination uncertainties since the widths of the synthesized beams vary significantly with declination
 (see \S\ref{sec:exposure}).\\
$^g$ Fluence completeness limits are given at the 90\% confidence level (see~\S\ref{subsec:completeness}). For sources observed twice a day, the second entry corresponds to the less sensitive lower transit.\\
$^h$ One of the three bursts had no intensity data captured; see Table~\ref{ta:bursts}.
\end{table}
\normalsize

\section{Analysis \& Results}
\label{sec:analysis}

We detected nine new repeating FRB sources with the CHIME/FRB instrument, using the criteria described in Section \ref{sec:obs}. Source properties are summarized in Table~\ref{ta:repeaters}, while individual burst properties are listed in Table \ref{ta:bursts}.

\subsection{Source localization}
\label{sec:localization}

\begin{figure}[b]
	\centering
\includegraphics[angle=0,scale=0.45]{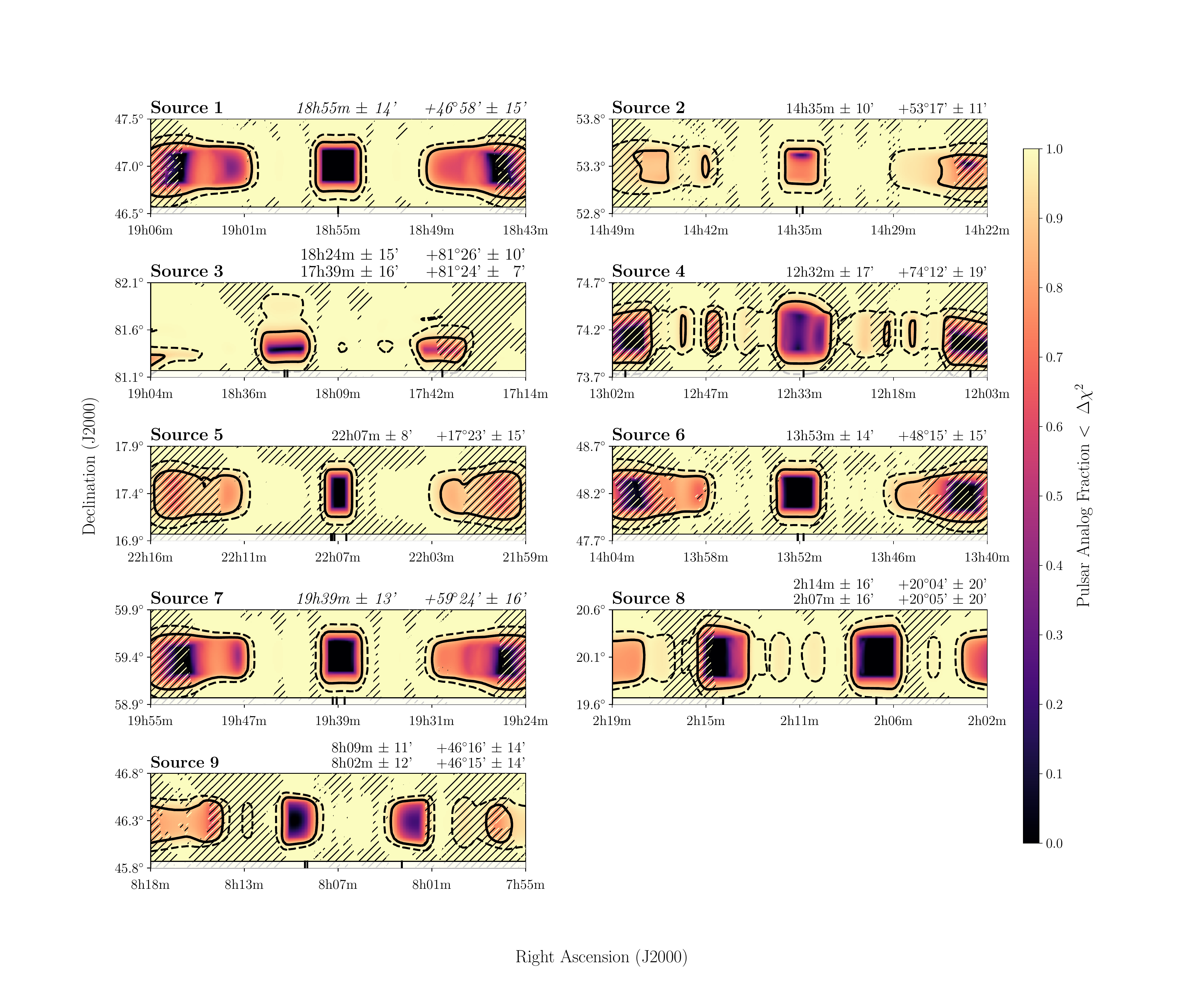}
\figcaption{Detection positions of the new CHIME/FRB repeating FRB sources, as determined from CHIME/FRB detection beam information through the methods described in \S\ref{sec:localization}. Each panel is $1^\circ \times 4^\circ$. localization is performed as a $\chi^2$-minimization. The method is applied to a large population of analogous pulsar events (i.e., pulsars with similar brightness and beam-detection statistics), which we use to translate $\Delta\chi^2$ values to empirical confidence intervals depicted by the color scale. The 90\% and 99\% confidence intervals are indicated as solid and dashed contours; we use the former interval to report the most likely positions. The R.A. of the beam centers for each detection are shown as black ticks on the bottom of each panel. For declination, panels are centered on the beam with the highest S/N detection; we do not add analogous ticks on the vertical axes since each beam's declination is constant in time. Hatched regions represent disfavored areas where, for at least one burst in the sample, the beam model predicts substantial attenuation in formed-beam sensitivity in the portion of the band where emission is observed.}
\label{fig:localization}
\end{figure}

Burst localization was carried out following the methods described in Paper I. For all sources, we used a model of the CHIME primary beam and S/N estimates from all beams that detected bursts in order to obtain burst-averaged sky positions using a $\chi^2$-grid method. While the methods remain identical to those used in Paper I, we have updated our underlying beam model to correct for an effective 0.071$^\circ$ rotation of the telescope relative to true North (counter clockwise as viewed from the CHIME meridian) that was identified after publication of Paper I.\footnote{This amount of rotation corresponds to a distance offset between the apparent and true N-S directions of $\sim$5 cm at the North end of each cylinder} We chose not to correct positions reported in Paper I in this work since their quoted uncertainties make the corrected positions statistically equivalent. Our best estimates of source positions are provided in Table~\ref{ta:repeaters}, and graphical depictions of the localization regions are shown in Figure~\ref{fig:localization}.  

We were not able to acquire total-intensity or baseband data for one burst from Source 3, since data acquisition was temporarily disabled around the time of this burst due to commissioning-related upgrades of the CHIME/FRB system. However, since this burst satisfied detection-S/N and classification thresholds for FRB signals, we nonetheless recorded the metadata produced by the realtime detection pipeline. Since these data contain the quantities needed for burst-averaged localization refinement, we used the metadata for this burst in generating $\chi^2$ grids and confidence intervals for the sky localization of Source 3.

\subsection{Exposure Determination}
\label{sec:exposure}
On-sky exposure (referenced to 600 MHz) across the FWHM region of the total-intensity beams synthesised for the CHIME/FRB system was determined in a manner similar to that used in Paper I. In summary, we estimated the exposure for a grid of sky positions within the 90\% confidence uncertainty region for each source (see Figure \ref{fig:localization}). We then calculated the weighted average and standard deviation for the exposure over all these positions, with the weights equal to the sky-position probability maps shown in Figure \ref{fig:localization}. Exposure for each source during the interval from 28 August 2018 to 30 September 2019 is reported in Table \ref{ta:repeaters} and plotted in Appendix \ref{app:exposure} (see Figure \ref{fig:exposure}). For high declination sources ($\delta > +70^\circ$), which transit across the primary beam of the telescope twice each day, we chose to report the exposure for the lower transit separately since the beam response is different for the two transits. 

\subsection{Determination of Burst Fluence and Peak Flux Density}
\label{sec:fluence}
Dynamic spectra were calibrated and burst fluences and peak fluxes were determined as described in Paper I. In summary, we used transit observations of steady sources with known spectral properties to obtain flux conversion factors as a function of frequency in the vicinity of each burst. These calibration spectra were then applied to the total-intensity burst data to achieve calibrated dynamic spectra in flux units corrected for sensitivity variations across our band due to the telescope primary beam. 

Fluences were calculated by integrating the extent of each burst in the band-averaged time series (binned at the full $0.98304\,\text{ms}$ resolution of total-intensity data) while peak fluxes were taken to be the highest value within the same extent. Due to the narrow-band nature of the bursts in our sample, averaging the signal over the entire bandwidth also averages noise into our fluence and flux values. However, we chose to quote burst fluence and fluxes from a the same frequency range for consistency. If there were multiple sub-bursts in a given burst, then peak flux and fluence values were obtained for each component. To simplify these calculations, we assumed that all bursts were detected along the meridian of the primary beam so that our fluences and fluxes represent lower bounds. Uncertainties were estimated by taking into account beam and time variations in system sensitivity using steady calibrator sources as in Paper I.

For all sources except Source 4, we used steady sources within $5^{\circ}$ of declination for calibration. Since there are no calibration sources within $5^{\circ}$ in declination from Source 4, we assumed that our beam is North-South symmetric and used a source within $1^{\circ}$ of declination on the opposite side of zenith.

\subsection{Fluence Completeness Determination}
\label{subsec:completeness}

We determined fluence completeness, the per-source threshold fluence for which all bursts above it are expected to be detected by the CHIME/FRB instrument, in a manner similar to that used in Paper I, with some minor modifications to how sensitivity variation is characterised. 

Previously, for circumpolar sources observed twice a day, fluence thresholds were simulated for the transit where bursts were detected, then calibrators with similar declination were used to get a band-averaged sensitivity scaling for the transit without detections. We caution that the per-frequency scaling between transits is highly structured, so the observed attenuation for a given source will depend strongly on its spectrum. In this work, we extended the simulations to cover both transits, using our beam model to compute relative sensitivities. This approach included a description of the synthesised beams in addition to the degradation in the primary beam sensitivity, which is significant for sources that pass between synthesised beams in one transit while crossing beam centers in the other transit.

In order to characterise the inter-day sensitivity variation of the CHIME/FRB system, we used the method described in \citet{jcf+19}. The method involves estimating the daily variation in RMS noise at the location of each source by analysing distributions of S/N values of pulsars within 5$^\circ$ of the source declination. However, observations with the CHIME/FRB system through 2019 June suggest that small variations in RMS noise can have a significant effect on the FRB detection rate.  Since these variations cannot be adequately characterised by the small number of pulsars detectable in a 5$^\circ$ declination range, we used pulsars which are robustly detected by the CHIME/FRB system within its entire observable declination range (with Dec. $> -11^\circ$) in this work. The estimate of the daily variation in RMS noise is obtained by averaging measurements from pulsars detected on each sidereal day for which the telescope was operating with the same gain calibration. This approach is in contrast to the method used in \citet{jcf+19}, which combined measurements from pulsars detected on the same UTC day. Our modification makes the measurement of the relative RMS noise more sensitive to changes in system sensitivity due to varying gain-calibration strategies. Typical daily variations are at the 20\% level.

\subsection{Characterisation of Burst Morphology}
\label{subsec:burstmorph}

\begin{figure}[t]
	\centering
\includegraphics[width=0.95\textwidth]{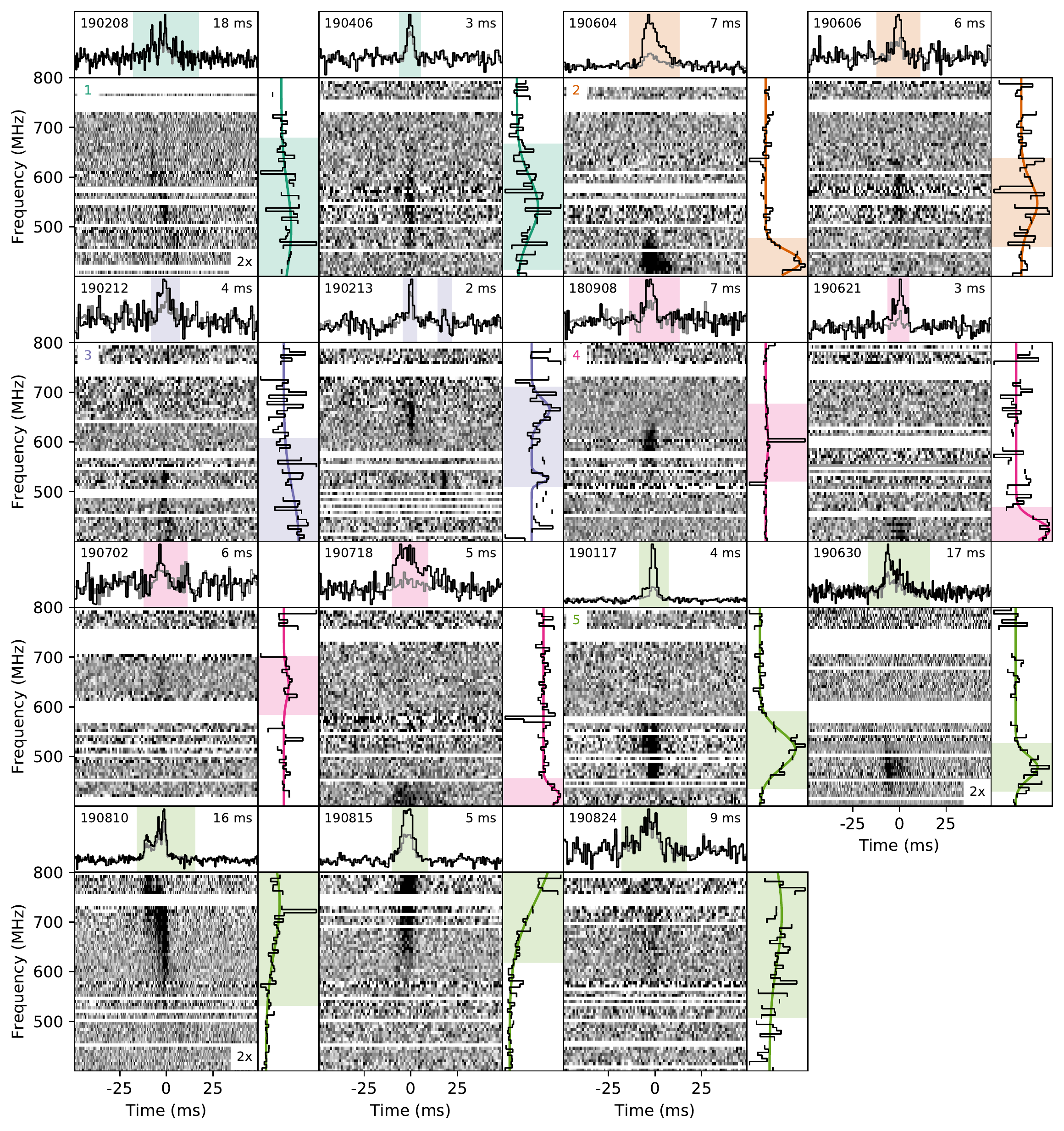}
\figcaption{Dynamic spectra of the bursts listed in Table~\ref{ta:bursts}, for the per-burst optimal DMs as determined in \S\ref{subsec:burstmorph}. Every panel shows the 0.98304-ms time resolution dedispersed intensity data with the integrated burst profile on top and the on-pulse spectrum on the right. Subsequent sources are colored differently, with the source number in the top left corner of each first burst. Windows show 100 time samples ($\sim$100 ms), unless indicated otherwise by the multiplicative factor in the bottom right corner. Intensity values are saturated at the 5$^\mathrm{th}$ and 95$^\mathrm{th}$ percentiles. All bursts were detected in the source's upper transit. Pulse widths, defined as the width of the boxcar with the highest S/N after convolution with burst profile, are in the top right corner. The shaded region in the profile (four times the pulse width) was used for the extraction of the on-pulse spectrum. The shaded region in the on-pulse spectrum shows the full width at tenth maximum (FWTM) of a Gaussian fit. In the burst profiles, the black lines are the integration over the FWTMs and the gray lines are the integration over the full bandwidths. 64 frequency subbands with a 6.25 MHz subband bandwidth are shown for all bursts. There are underlying missing or masked channels at the full-resolution (16,384-frequency-channel) intensity data, resulting in an average effective bandwidth of 214 MHz.} %Underlying missing or masked channels of the full-resolution (16,384-frequency-channel) intensity data are depicted by red lines on the left of the intensity data.}
\label{fig:waterfall1}
\end{figure}

\begin{figure}[t]
	\centering
\includegraphics[width=0.95\textwidth]{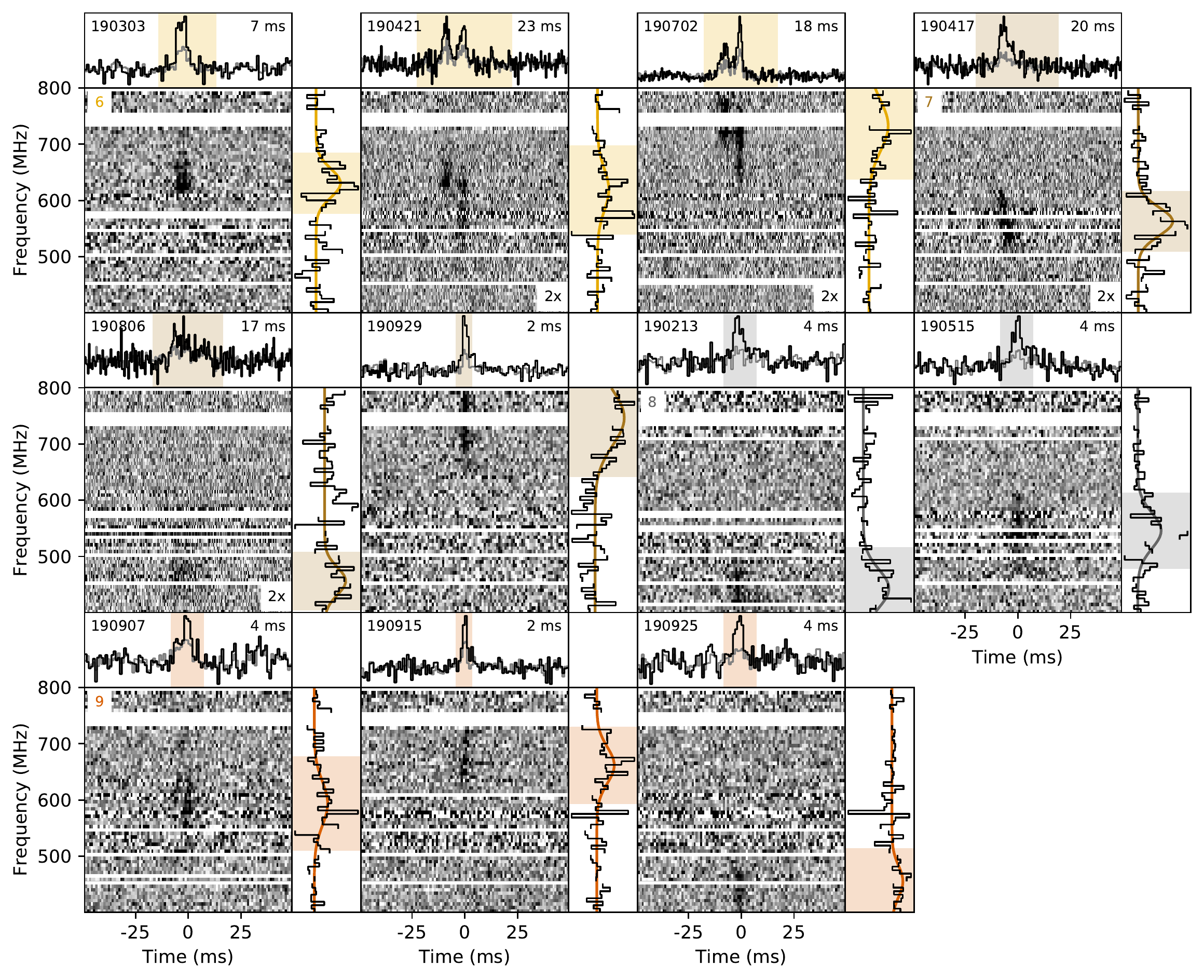}
\figcaption{Same as Fig.~\ref{fig:waterfall1}.}
\label{fig:waterfall2}
\end{figure}

Many bursts from repeating FRB sources exhibit complex morphology, comprised of multiple sub-bursts that usually drift down in frequency as time progresses \citep[e.g.,][]{hss+19}. Finding the optimal DM necessitates methods beyond the typically employed S/N-optimisation, and here we used the same method as in Paper I --- maximising burst structure by calculating the phase coherence of emission in all frequency channels with the \texttt{DM\_phase} package\footnote{\url{https://github.com/danielemichilli/DM_phase}} (Seymour et al. in prep.) over a range of trial DMs. The sub-burst alignment after dedispersion was verified by eye and Figures~\ref{fig:waterfall1} and \ref{fig:waterfall2} show dynamic spectra for all bursts, dedispersed to the best estimate of that burst's DM, as listed in Table~\ref{ta:bursts}.

We used the same modeling procedure discussed in Paper I for estimating widths, arrival times and scattering timescales from our calibrated total-intensity dynamic spectra, which we summarize here. In all fits, we preserved the raw time and frequency resolution of the CHIME/FRB total-intensity data and fitted two-dimensional models of Gaussian temporal profiles and either Gaussian or weighted power-law spectral shapes. For bursts with simple morphology i.e., with no significant multiple components), we fitted a single-burst spectrum and allowed the DM to float freely during each fit in order to determine robust uncertainties for all other burst-specific parameters. In these cases, the DMs estimated from direct modeling of the dynamic spectra were statistically consistent with the values estimated from the structure-optimisation algorithm shown in Table \ref{ta:bursts}. For bursts with spectro-temporal structure typical of repeating sources, we fitted several components and held the DM fixed to the structure-optimisation value in Table \ref{ta:bursts} in order to ensure separation of sub-burst components and the determination of robust width estimates. 

For all bursts, we also fitted single-tail scattering profiles to dynamic spectra that arise from frequency-dependent, multi-path propagation of the signal through small-scale electron density inhomogeneities in the ISM. We fitted for scattering by applying a temporal pulse broadening function \cite[e.g.][]{mck14} and assuming that the scattering timescales depends on frequency as $f^{-4}$. As done in previous works, we selected the superior model for each burst -- one that explicitly fits for the scattering timescale or one that does not fit for scattering effects -- based on the best-fit $\chi^2$ statistic \citep[see ][and Paper I]{abb+19a}. Using this model selection procedure, we consider all widths reported in Table \ref{ta:bursts} to approximate the intrinsic burst widths. Significant estimates of the scattering timescale are presented with uncertainties in Table \ref{ta:bursts}; for non-detections of scattering timescales and widths, which we defined to be consistent with 0 ms at 3$\sigma$ confidence, we present upper limits of the 2$\sigma$ confidence interval. As noted in Paper I, repeater morphology likely introduces bias into fits of scattering timescales due to faint, extended emission being detected as part of a scattering tail; we therefore urge caution when interpreting scattering timescales shown in Table \ref{ta:bursts}.

As in Paper I, we determined burst drift rates using an autocorrelation analysis and present results for bursts where the drift rate is constrained in Table \ref{ta:bursts}. We calculated each linear drift rate by fitting a two-dimensional Gaussian profile to the two-dimensional auto-correlation. We used a Monte Carlo method to obtain robust confidence intervals, de-dispersing our bursts to 100 DM values drawn from the DM uncertainty distribution and fitting a linear drift rate for each of 100 random noise realisations per DM value. We found five bursts with significant drift rates (see Table \ref{ta:bursts}).

\begin{table}[t]
\begin{center}
\caption{Individual Burst Properties from Nine New CHIME/FRB Repeaters.$^a$ }
\hspace{-1.in}
\resizebox{1.1\textwidth}{!}{ 
\begin{tabular}{lclcccccc} \hline
    Day & MJD & Arrival Time$^b$  &  DM & Drift Rate & Width$^b$ & Scattering Time & Fluence$^f$ & Peak Flux Density$^f$ \\
  (yymmdd) &     & (UTC @ 600 MHz)  & (pc~cm$^{-3}$) & (MHz/ms) & (ms) & (ms @ 600 MHz)& (Jy ms) & (Jy) \\\hline
        \multicolumn{9}{c}{Source 1 (FRB 190208.J1855+46)} \\\hline
 190208 & 58522 & 17:41:42.300(3) & 579.9(2) & $-9.1_{-0.7}^{+0.9}$ & 0.91(16) / 0.62(15) / 0.84(15) / 1.2(3) & $<1.8$ & 1.4(6) / 0.8(3) / 1.1(5) / 0.6(3) & 0.4(2) / 0.4(3) / 0.6(3) / 0.3(2) \\
 190406 & 58579 & 13:55:15.0416(13) & 580.2(2) & ... & 1.31(14) & $<1.6$ & 2.0(8) & 0.6(3) \\\hline
 \multicolumn{9}{c}{Source 2 (FRB 190604.J1435+53)} 
 \\\hline
 190604 & 58638 & 05:49:49.2554(11) & 552.6(2) & ... & 3.0(4) & 1.7(4) & 8.3(2.8) & 0.9(4) \\
 190606$^c$ & 58640 & 05:34:23.574(3) & 552.7(2) & ... & 1.2(5) & $<2.2$ & 1.7(7) & 0.6(3) \\\hline
 \multicolumn{9}{c}{Source 3 (FRB 190212.J18+81)} \\\hline
 190212 & 58526 & 16:17:19.174(5) & 301.7(3) & ... & 4.1(1.6) & $<4.1$ & 3.0(1.5) & 0.4(3) \\
 190213 & 58527 & 17:09:28.419(3) & 301.4(2) & ... & 2.1(3) / 0.69(16) & 0.4(1) & 2.5(1.0) / 1.0(5) & 1.1(6) / 0.6(4) \\
 190516$^e$ & 58619 & ... & ... & ... & ... & ... & .. & ...\\\hline
 \multicolumn{9}{c}{Source 4 (FRB 180908.J1232+74)} \\\hline
 180908 & 58004 & 21:13:01.2578(4) & 195.7(9) & ... & 1.91(10) & $<2.1$ & 2.7(1.1) & 0.6(4) \\
 190621$^{c, d}$ & 58655 & 02:21:21.1131(15) & 195.7(9)$^d$ & ... & 2.8(3) & $<3.1$ & 1.1(3) & 0.4(2) \\
 190702 & 58666 & 01:58:11.8666(6) & 195.4(4) & ... & 1.6(3) & $<2.2$ & 1.4(6) & 0.5(4) \\
 190718$^{d}$ & 58682 & 01:11:20.056(3) & 196.1(8)$^d$ & ... & 9(2) & 1.29(3) & 2.9(1.8) & ... \\\hline
 \multicolumn{9}{c}{Source 5 (FRB 190117.J2207+17)} \\\hline
 190117 & 58500 & 22:18:26.46617(7) & 393.3(1) & ... & 1.44(3) & $<1.5$ & 5.9(1.6) & 1.7(6) \\
 190630$^c$ & 58664 & 11:34:38.8739(4) & 392.6(6) & ... & $<1.3$ & 5.0(5) & 6.7(1.3) & 0.6(2) \\
 190810 & 58705 & 08:52:24.060(6) & 392.9(2) & $-12.1_{-1}^{+0.9}$ & 3.2(3) / 3.9(4) / 0.64(15) & 3.7(3) & 5.0(1.4) / 7.1(1.6) / 12(2) & 0.7(3) / 1.7(5) / 2.6(6) \\
 190815 & 58710 & 08:33:46.5997(5) & 395.1(2) & ... & 2.56(13) & 1.9(7) & 9.1(2.2) & 1.4(4) \\
 190824 & 58719 & 07:58:55.880(3) & 396.5(7) & ... & 5.2(1.1) & $<7.4$ & 5.1(1.1) & 0.6(3) \\\hline
   \multicolumn{9}{c}{Source 6 (FRB 190303.J1353+48)} \\\hline
 190303 & 58545 & 11:05:13.695(1) & 221.8(5) & ... & 2.0(3) & 1.8(7) & 2.3(9) & 0.5(3) \\
 190421 & 58594 & 08:00:07.2616(6) & 223.5(3) & $-12_{-3.6}^{+1.4}$ & 3.1(3) / 3.5(5) & $<4.5$ & 2.0(5) / 2.3(6) & 0.4(2) / 0.4(2) \\
 190702$^c$ & 58666 & 03:14:39.7628(3) & 222.4(2) & $-11.02_{-1.2}^{+0.9}$ & 4.5(3) / 2.0(1) & $<5.1$ & 3.1(8) / 3.1(8) & 0.5(2) / 1.0(4) \\\hline
   \multicolumn{9}{c}{Source 7 (FRB 190417.J1939+59)} \\\hline
 190417 & 58590 & 13:55:44.814(3) & 1378.1(2) & $-7.06_{-0.5}^{+0.3}$ & 3.3(9) & 3.1(1.1) & 4.4(8) & 0.5(2) \\
 190806 & 58701 & 06:36:49.674(6) & 1379(1) & ... & 9(3) & $<15$ & 3.2(7) & 0.4(2) \\
 190929 & 58755 & 03:14:15.846(2) & 1378.5(3) & ... & 1.19(2) & $<1.3$ & 1.7(4) & 0.7(2) \\\hline
    \multicolumn{9}{c}{Source 8 (FRB 190213.J02+20)} \\\hline
 190213 & 58527 & 00:42:17.295(4) & 651.1(4) & ... & 10(2) & $<2.2$ & 0.6(3) & ... \\
 190515 & 58618 & 18:33:37.853(2) & 651.5(4) & ... & $<4$ & 4.0(1.1) & 3.0(1.2) & 0.5(3) \\\hline
    \multicolumn{9}{c}{Source 9 (FRB 190907.J08+46)} \\\hline
 190907 & 58733 & 17:02:43.311(3) & 310.0(4) & ... & 3(1) & $<5$ & 1.7(6) & 0.3(2) \\
 190915 & 58741 & 16:27:36.8563(4) & 309.5(3) & ... & 0.54(14) & 1.7(4) & 0.9(4) & 0.4(2) \\
 190925 & 58751 & 15:54:06.438(2) & 309.5(2) & ... & 3.0(8) & $<4.6$ & 0.7(2) & 0.2(1) \\\hline
\end{tabular}}
\label{ta:bursts}
\end{center}
$^a$ Unconstrained parameters are listed as ``...''  Uncertainties are reported at the $1\sigma$ confidence level. Reported upper limits are those of the $2\sigma$ confidence level.\\
$^b$ All burst times of arrival are topocentric. Bursts with multiple components have one topocentric arrival time and several widths, fluences, and peak flux densities reported; the arrival time refers to the first sub-burst, and width, fluence, and peak flux density values for each component are presented in order of arrival.\\
$^c$ Baseband data recorded for the burst.\\
$^d$ From S/N-optimization. \\
$^e$ No total-intensity or baseband data were recorded for this event. See Section \ref{sec:localization} for details. \\
$^f$ Fluence and peak-flux-density measurements represent lower bounds as we assumed all bursts were detected along the meridian of the primary beam.
\end{table}

\subsection{Baseband Detections}
\label{sec:basebandR3}

The CHIME/FRB baseband system \citep[see][]{abb+18} was triggered for one burst each from Sources 2, 4, 5, and 6. We used the baseband data for these events to verify the localization regions obtained from the per-beam S/N analysis described in Section \ref{sec:localization}; the results and details of future position refinement with baseband data will be presented elsewhere (Michilli et al., in prep.). Once available, this analysis will reduce the localization precision to $\sim\frac{6.7}{S/N}$\,arcmin \citep{abb+18}. Visual inspection of the baseband dynamic spectra confirmed successful capture of all four bursts by the baseband system. All baseband events were incoherently dedispersed to the S/N-optimising DM after coherently dedispersing each of the 1,024 baseband channels to a nearby fiducial value, in order to mitigate intra-channel smearing. 

All four baseband events were manually processed through a polarization analysis pipeline. The pipeline, described in an upcoming paper (Mckinven et al., in prep.), searches for an RM detection using the {\tt RM-tools} package\footnote{\url{https://github.com/CIRADA-Tools/RM}} that implements two independent methods: RM-synthesis \citep{burn66,bd05} and Stokes QU-fitting \citep{obr+12}. Using these algorithms, we detected moderate RM values for Sources 2 and 6. 

Table~\ref{tab:RMs} summarises the main polarization products for sources with an RM detection. RMs derived from RM-synthesis ($\mathrm{RM_{FDF}}$) and Stokes QU-fitting ($\mathrm{RM_{QUfit}}$) are reported, and show small but significant differences. $\mathrm{RM_{FDF}}$ is determined from the peak of the Faraday dispersion function (FDF), while $\mathrm{RM_{QUfit}}$ is calculated from a parametric fit under a thin-screen model of Faraday rotation. Uncertainties in $\mathrm{RM_{FDF}}$ were determined in a manner consistent with previous analysis of a different burst presented in Paper I, while those for $\mathrm{RM_{QUfit}}$ were determined from its marginal probability density output by the Nested Sampling routine. The differences in estimates of $\mathrm{RM_{FDF}}$ and $\mathrm{RM_{QUfit}}$ are a more valid measure of the systematic RM measurement uncertainty than each method's formal measurement error. The discrepant RMs are also a reflection of small differences in the measurement procedure of the two methods. In particular, the Stokes Q,U fitting of {\tt RM-tools} implements a Nested Sampling algorithm \citep{ski04} to find the best-fitting parameters. This method allows the partially degenerate parameters, RM and the de-rotated polarization angle ($\chi_0$), to be simultaneously fit. 

%In addition, an estimate of linear polarized fractions is included. Source 6 appears to have a much lower value than the $\sim$100\% linear polarization fractions of other repeating FRBs, both reported here and elsewhere \citep[][Paper I]{michilli2018extreme}. Burst profiles for total intensity (I), linear (L) and circular polarization (V) are shown in Figure~\ref{fig:pol_profs} along with the polarization position angle after correcting for Faraday rotation using $\mathrm{RM_{QUfit}}$ values output by the Stokes QU-fitting implementation.

Figures~\ref{fig:baseband_source2} and \ref{fig:baseband_source6} summarise the general properties of the polarized signal for Sources 2 and 6, respectively: dynamic spectra for Stokes I, Q, U and V parameters where data have been downsampled to $\sim$1.56-MHz and $\sim$0.33-ms resolution; FDFs that are cleaned of instrumental response introduced by limited bandwidth coverage \citep{hea09}; and Stokes Q,U spectra normalised by the total linear polarization at each frequency. 

We note that mixing of polarized signal between different Stokes parameters was not corrected for and appears to affect sources to varying degrees. In particular, Source 6 (see Fig.~\ref{fig:baseband_source6}) displays significant leakage of signal of Stokes U into Stokes V, as evidenced by a $\lambda^2$ modulation consistent with Faraday rotation. This leakage makes the uncorrected linear polarized fraction (L/I, with $\mathrm{L = \sqrt{Q^2 + U^2}}$) for Source 6 a lower bound, which we estimate to be L/I $\ge$ 0.2, and the circular polarized component highly uncertain. Leakage appears to be subdominant for Source 2, as its polarized fraction is consistent with unity. Corrections of these instrumental effects are a work in progress and, although important for accurate linear and circular polarized fractions, should not substantially change the RM values reported here.   

\begin{figure}[t]
	\centering
    \includegraphics[width=0.4\textwidth]{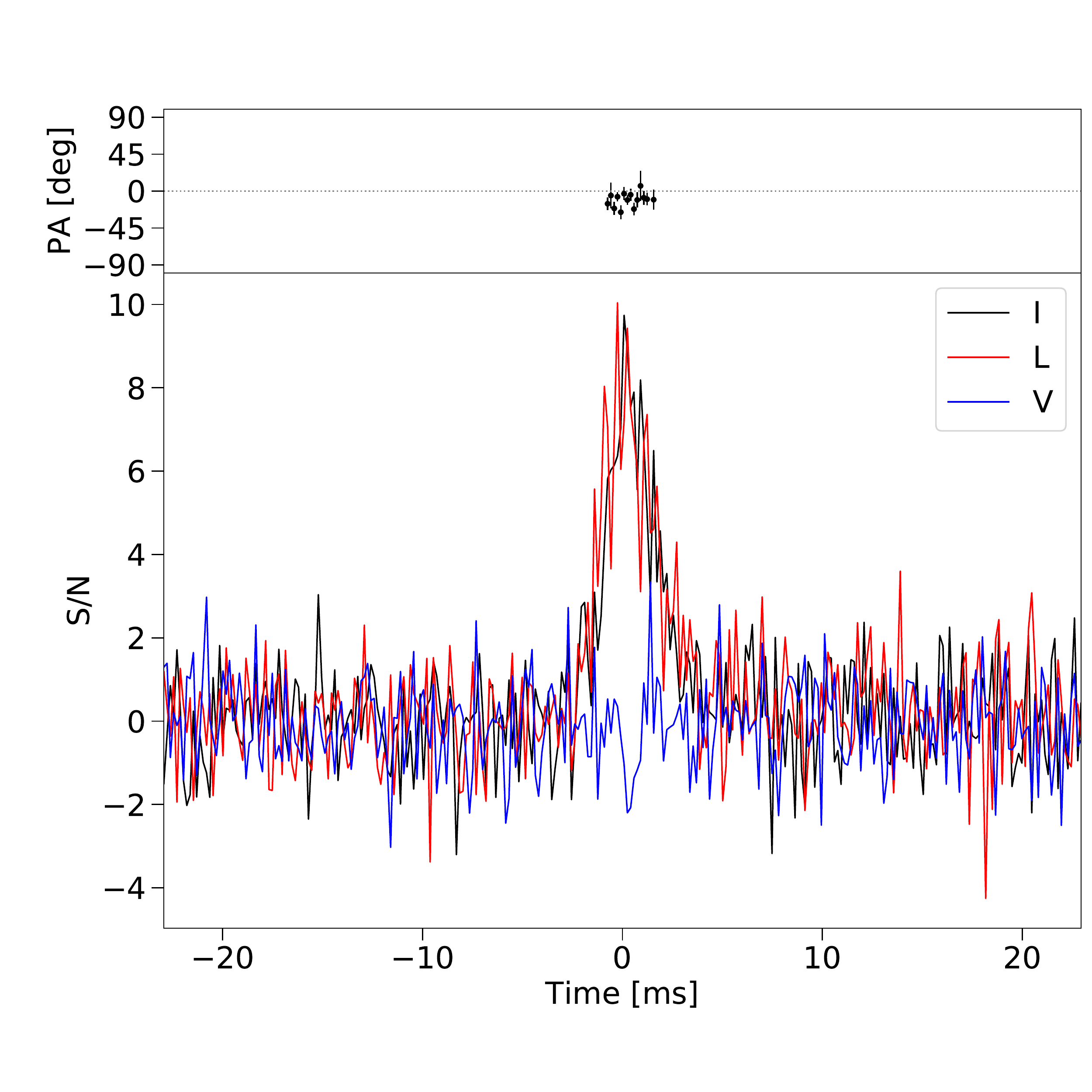}
    \includegraphics[width=0.4\textwidth]{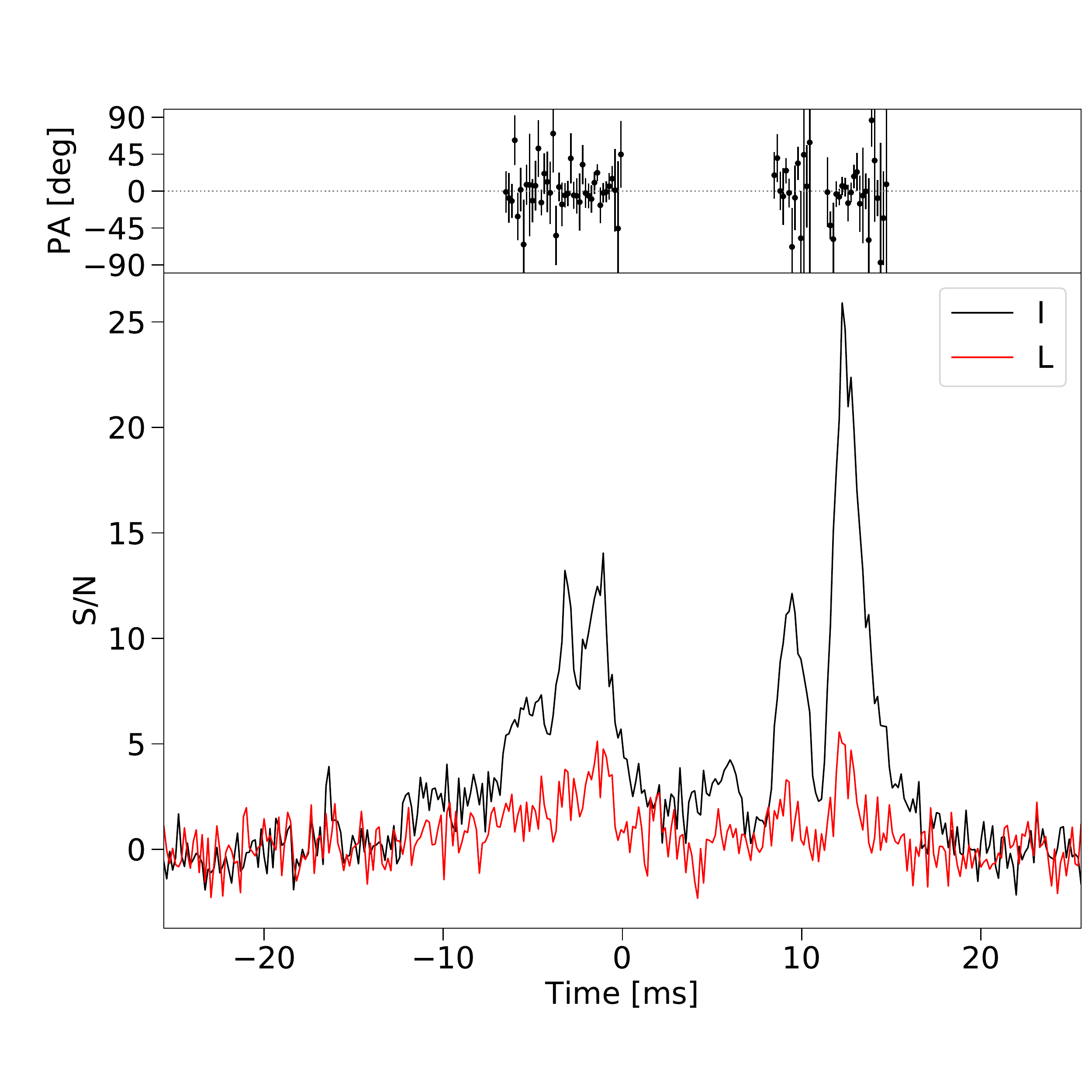}
    \figcaption{Pulse profiles (left to right: Source 2 and Source 6) for total intensity (I, black), linear polarization (L, red) after correcting for the detected RM, circular polarization (V, blue) and the uncalibrated polarization position angle (upper panel). For Source 6, the linear polarized burst profile is a lower limit while the the circular polarization profile has been omitted to avoid misleading conclusions drawn from data that has yet to be robustly corrected for instrumental leakage. See discussion in Section \ref{subsec:RMimplications} for more details.}
\label{fig:pol_profs}
\end{figure}

\begin{table}[h!]
%\centering
\begin{center}
\caption{Observed and Galactic-foreground RM values for CHIME/FRB repeaters.}
\begin{tabular}{ cccccc } 
 \hline
 Source & Day & $\mathrm{RM_{FDF}}$  & $\mathrm{RM_{QUfit}}$ &  $\mathrm{RM^{Oppermann+}_{MW}}^c$  & $\mathrm{RM^{GMIMS}_{MW}}^d$\\%& L/I \\ 
        & (yymmdd) & (rad m$^{-2}$) & (rad m$^{-2}$) & (rad m$^{-2}$) & (rad m$^{-2}$) \\
 \hline
 2 & 190606 & $-$16 $\pm$ 1 & $-$20 $\pm$ 1 & 13 $\pm$ 5 & $-$30 $\pm$ 4 \\%& $\sim$ 1 \\ 
 6 & 190702 & $-$504.4 $\pm$ 0.4 & $-$499.8 $\pm$ 0.7 & 14 $\pm$ 5 & $-$7 $\pm$ 10 \\%& $\gtrapprox$0.2 \\ 
 FRB 180916.J0158+65$^a$ & 181226$^b$ & $-$114.6 $\pm$ 0.6$^b$ & ...$^b$ & $-$72 $\pm$ 23 & $-$12 $\pm$ 6 \\
 \hline 
\end{tabular}
\end{center}
\label{tab:RMs}
$^a$ Source 1 in Paper I. \\
$^b$ Data taken from Paper I. \\
$^c$ Derived from the model developed by \cite{oppermann2015estimating}.\\
$^d$ Derived from the model developed by \cite{dlt+19}.
\end{table}

\begin{figure}[t]
	%\centering
\includegraphics[width=0.45\textwidth]{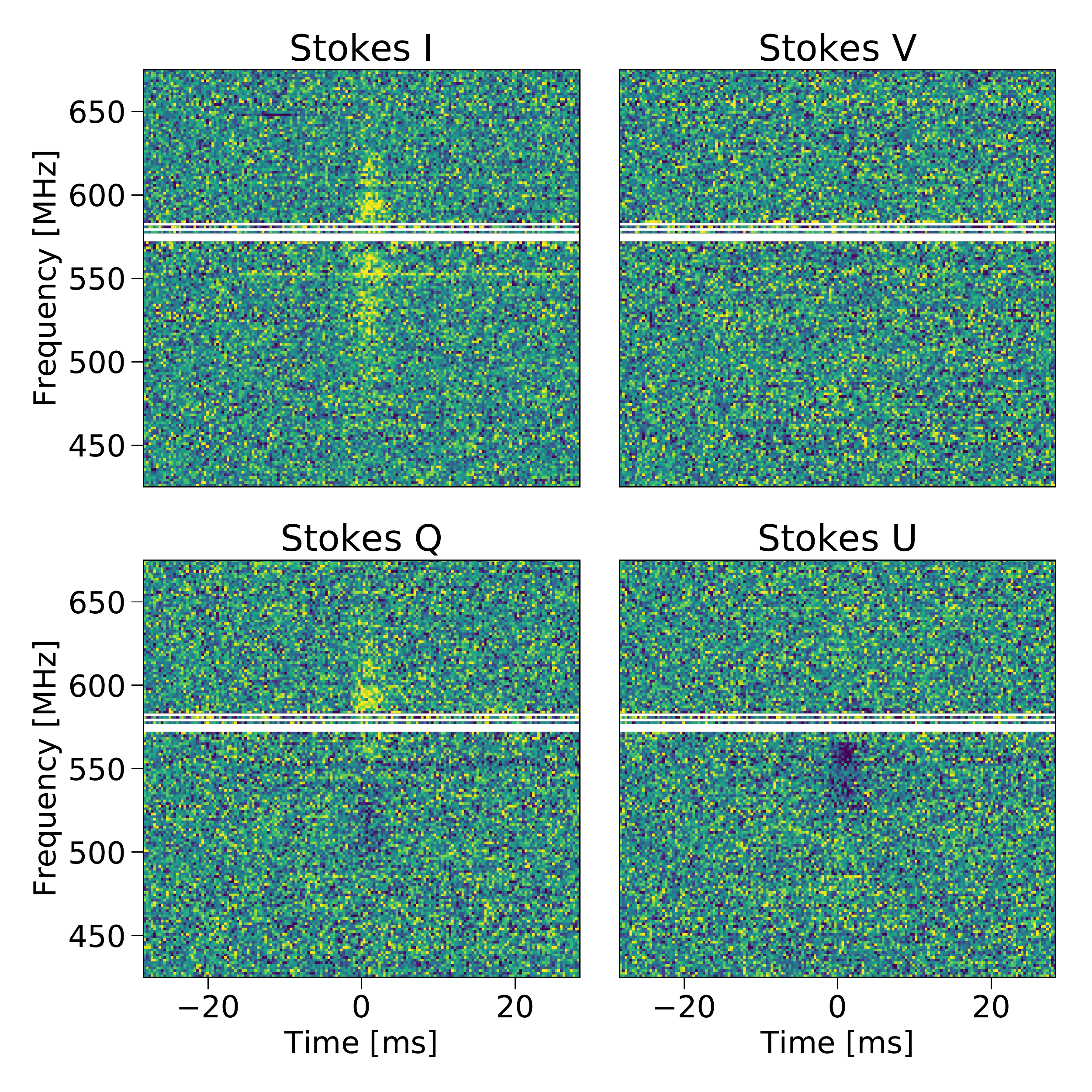}
\includegraphics[width=0.45\textwidth]{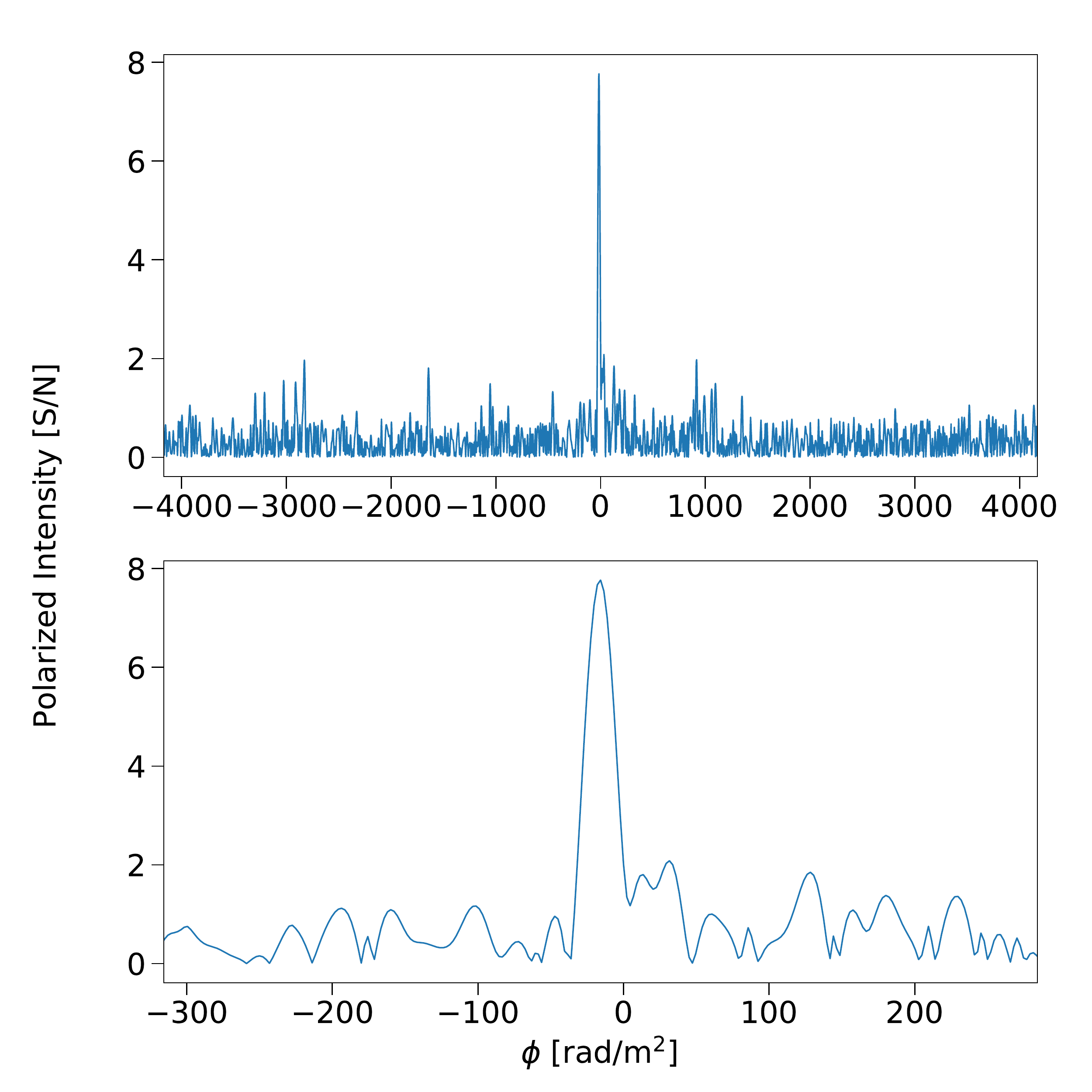}\\
\includegraphics[width=0.9\textwidth]{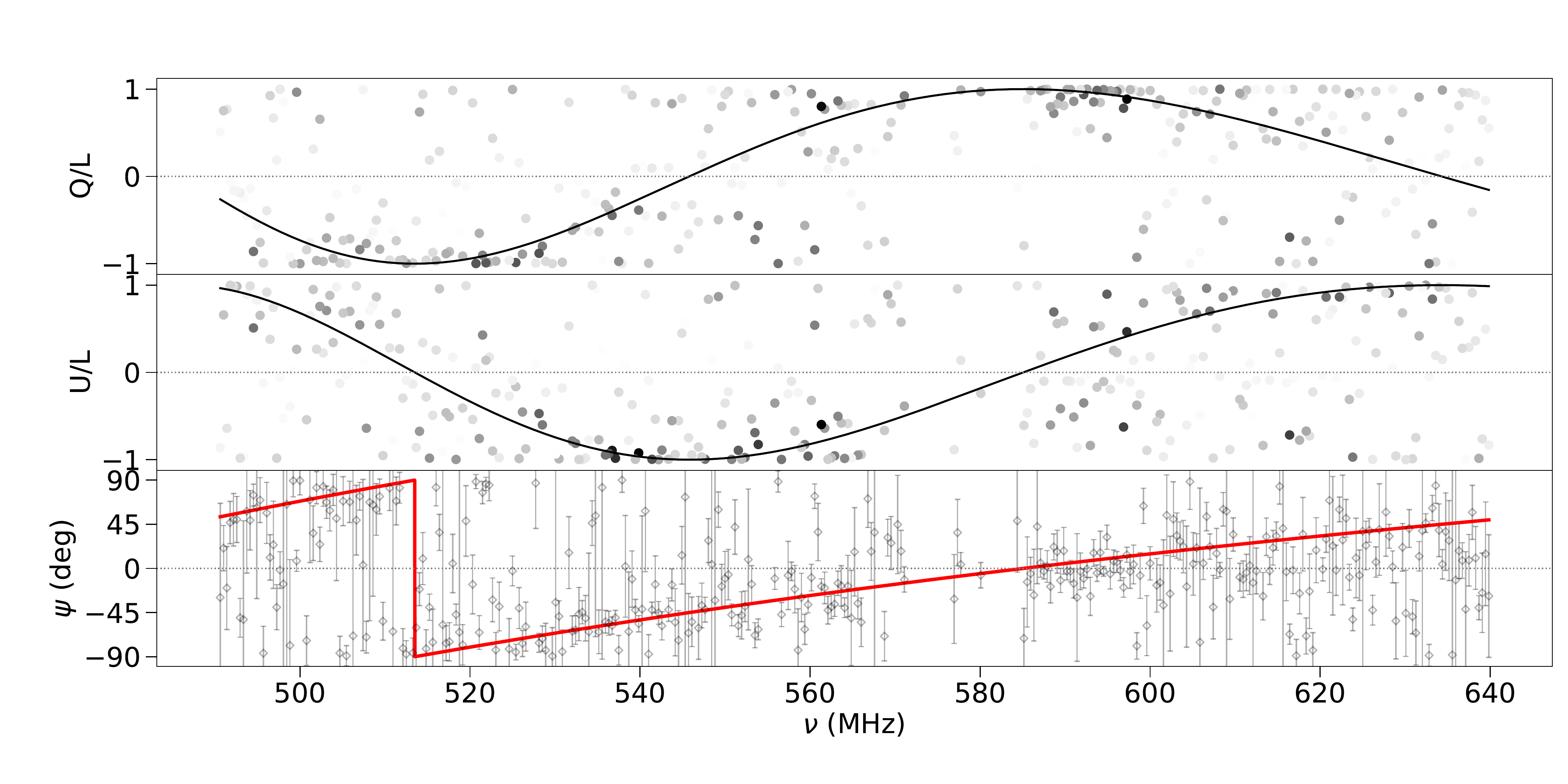}
\figcaption{Baseband data of the 190606 burst of Source 2.
    Top, left: Dedispersed dynamic spectra of Stokes I, Q, U \& V Stokes parameters.
    Top, right: The cleaned FDF showing polarized signal in S/N units out to a maximum Faraday depth determined by a 50\% drop in polarized sensitivity (upper panel) and the equivalent plot constrained over the region surrounding the peak of the FDF (lower panel).
    Bottom: Modulation of Stokes Q and U parameters with frequency, normalised by the total linear polarization (L). Frequency channels with significantly polarized signal are highlighted through a greyscale that saturates at higher S/N. The black curve is the best-fit model after applying a parabolic fit to the Stokes I spectrum and modulating the Stokes Q, U using the best-fit Faraday rotation. The lower panel shows the variation in the {\ef uncalibrated} polarization angle ($\psi$) as a function of frequency along with the best-fit model as a red line.
}
\label{fig:baseband_source2}
\end{figure}

\begin{figure}[t]
	%\centering
\includegraphics[width=0.45\textwidth]{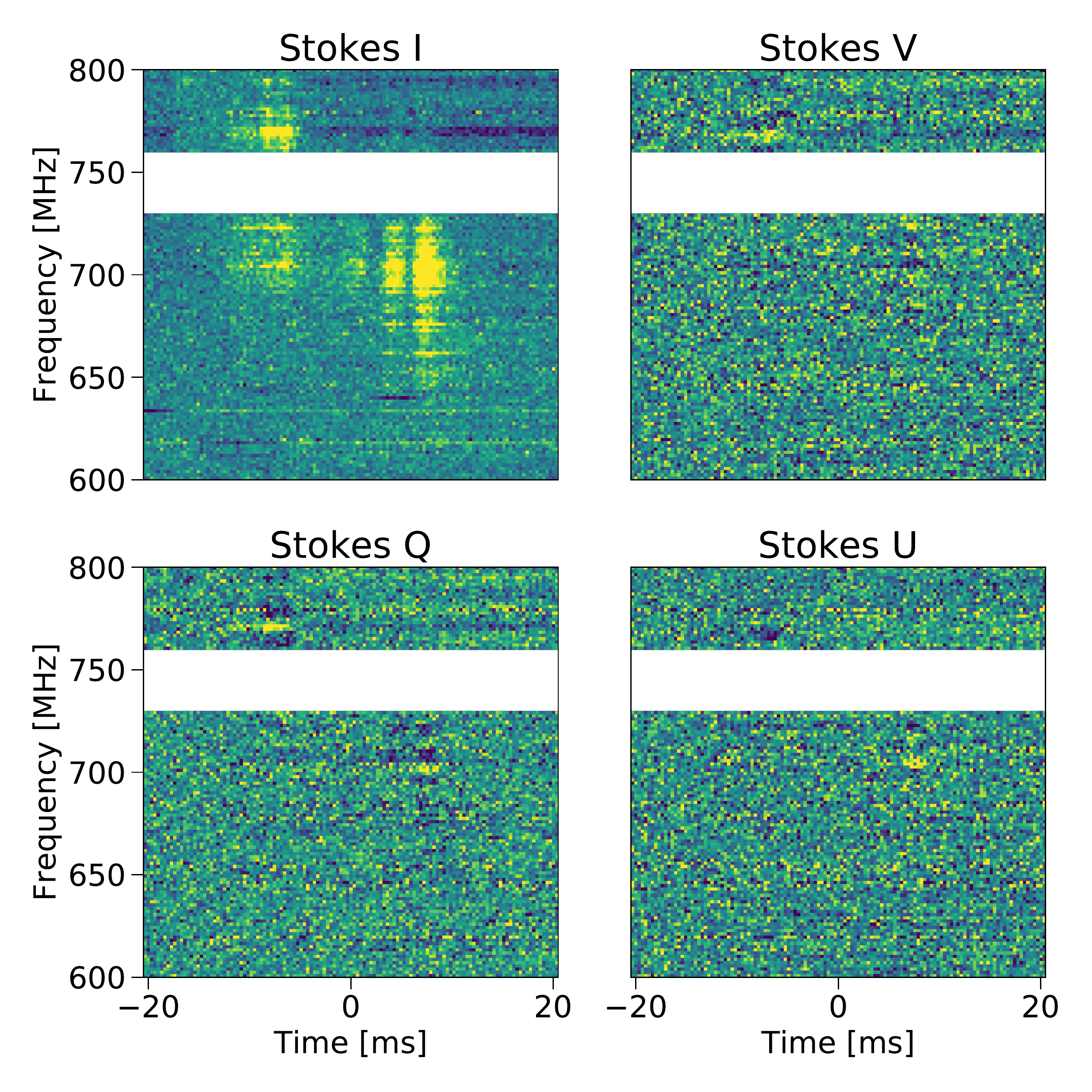}
\includegraphics[width=0.45\textwidth]{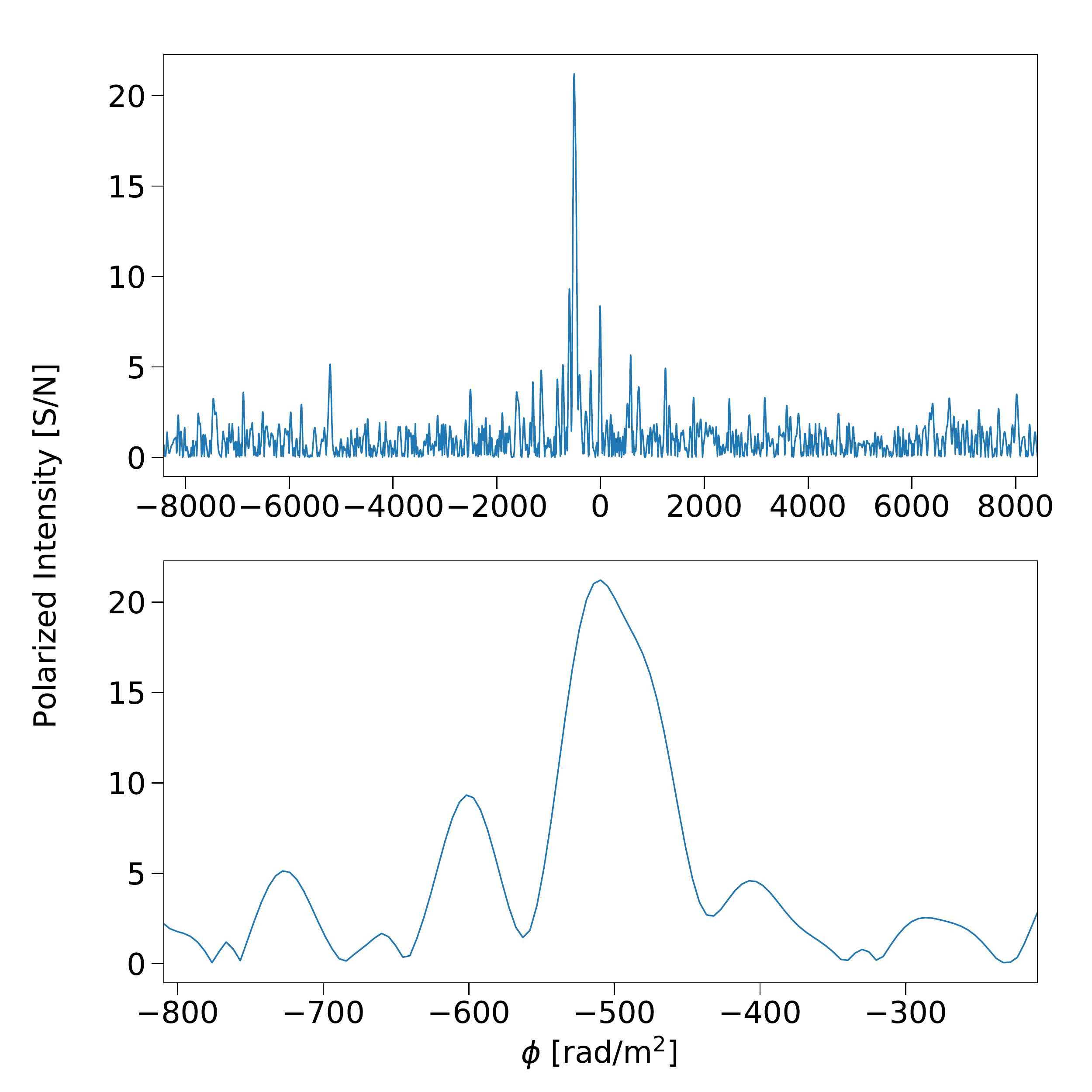}\\
\includegraphics[width=0.9\textwidth]{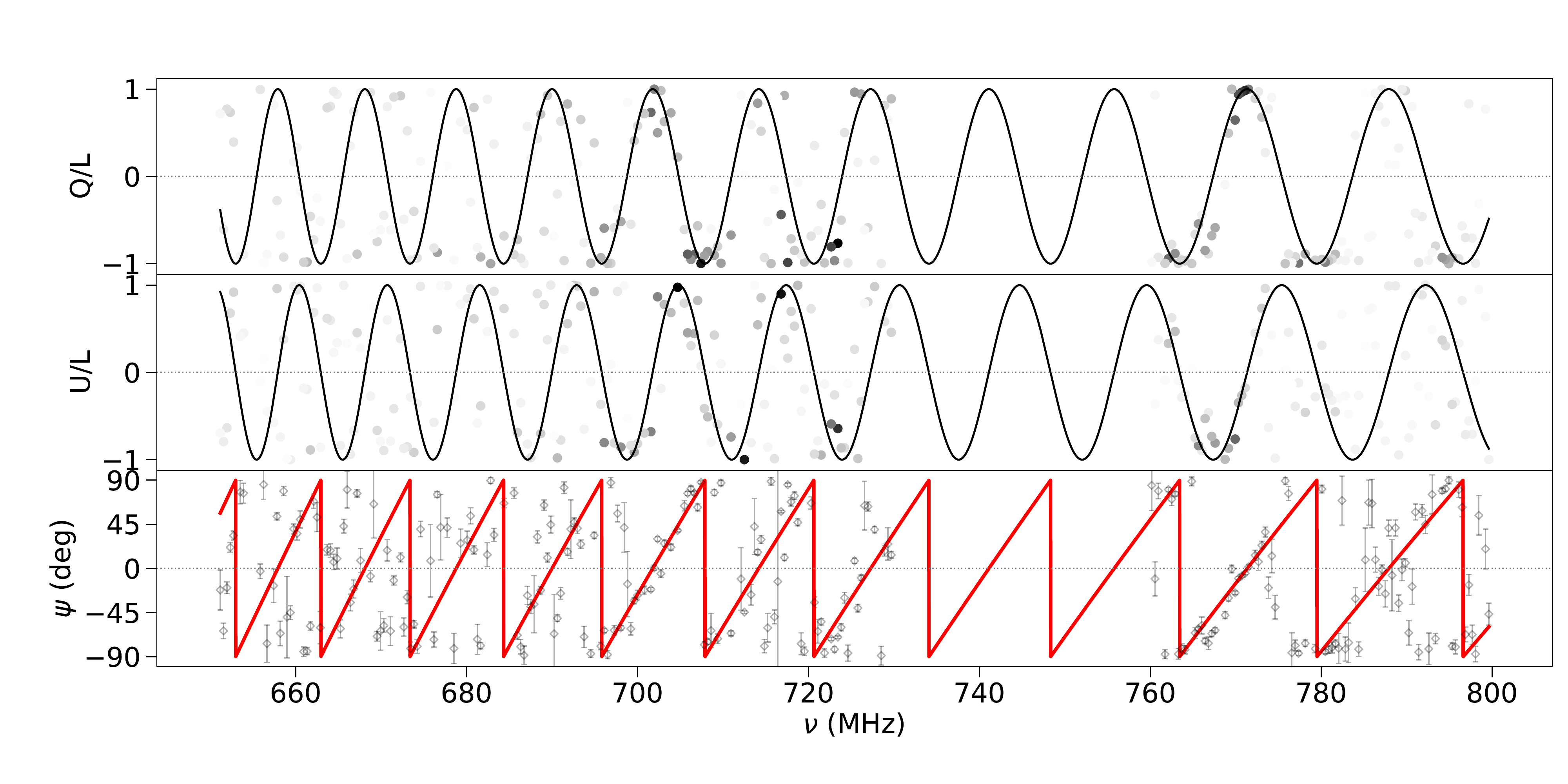}
\figcaption{Same plots as those in Figure~\ref{fig:baseband_source2} for baseband data of the 190702 burst of Source 6.}
\label{fig:baseband_source6}
\end{figure}

\section{Discussion}
\label{sec:discussion}

The systematic monitoring of a large sky fraction by CHIME/FRB has enabled the discovery of nine new repeating FRB sources, and a total of 18 repeating sources when including the results from \citet{abb+19b} and Paper I. {\ef The current sample of known repeating FRB sources now consists of 20 members, with FRBs 121102 \citep{ssh+16b} and 171019 \citep{kso+19,pat19} discovered using other observatories.} Besides offering opportunities for interferometric localization and multi-wavelength follow up \citep[e.g.,][]{mnh+20}, this number is large enough to enable studies of distributions of some source properties, although caution is needed as the CHIME/FRB pipeline has detection biases that remain to be quantified.

\subsection{Burst DMs and Morphologies}
\label{subsec:burstmorphdiscussion}

We compared the distribution of DMs of the 18 CHIME/FRB repeaters with that of the {\ef 12} published thus far non-repeating sources found by CHIME/FRB, and found no statistically significant difference using either a Kolmogorov-Smirnov or Anderson-Darling test.  Both distributions are subject to similar (though not necessarily identical) detection biases. This fact is suggestive that both their distribution in space and their associated local environments do not differ strongly, unless the differences cancel one another, which seems unlikely. Further comparison with additional CHIME/FRB sources that have not yet repeated (currently under analysis) will also be of interest.

Paper I noted a statistically significant difference in repeater and thus far non-repeater burst widths.  We repeated this analysis for this work with the new nine repeating sources included, using the $2\sigma$ confidence upper limits where there are no significant measurements of the pulse width. {\ef As discussed in Section \ref{subsec:burstmorph}, two best-fit models were obtained for each burst in the CHIME/FRB sample -- one that directly models the effect of scattering and another that ignores scattering -- and the model that yielded a better goodness of fit statistic was chosen as the superior model. Therefore, all widths were obtained in a similar manner and likely reflect the intrinsic burst widths when accounting for effects from scattering and multiple burst components that is typical in repeating-FRB activity.}

\begin{figure}[h]
	\begin{center}
		\includegraphics[width=0.85\textwidth]{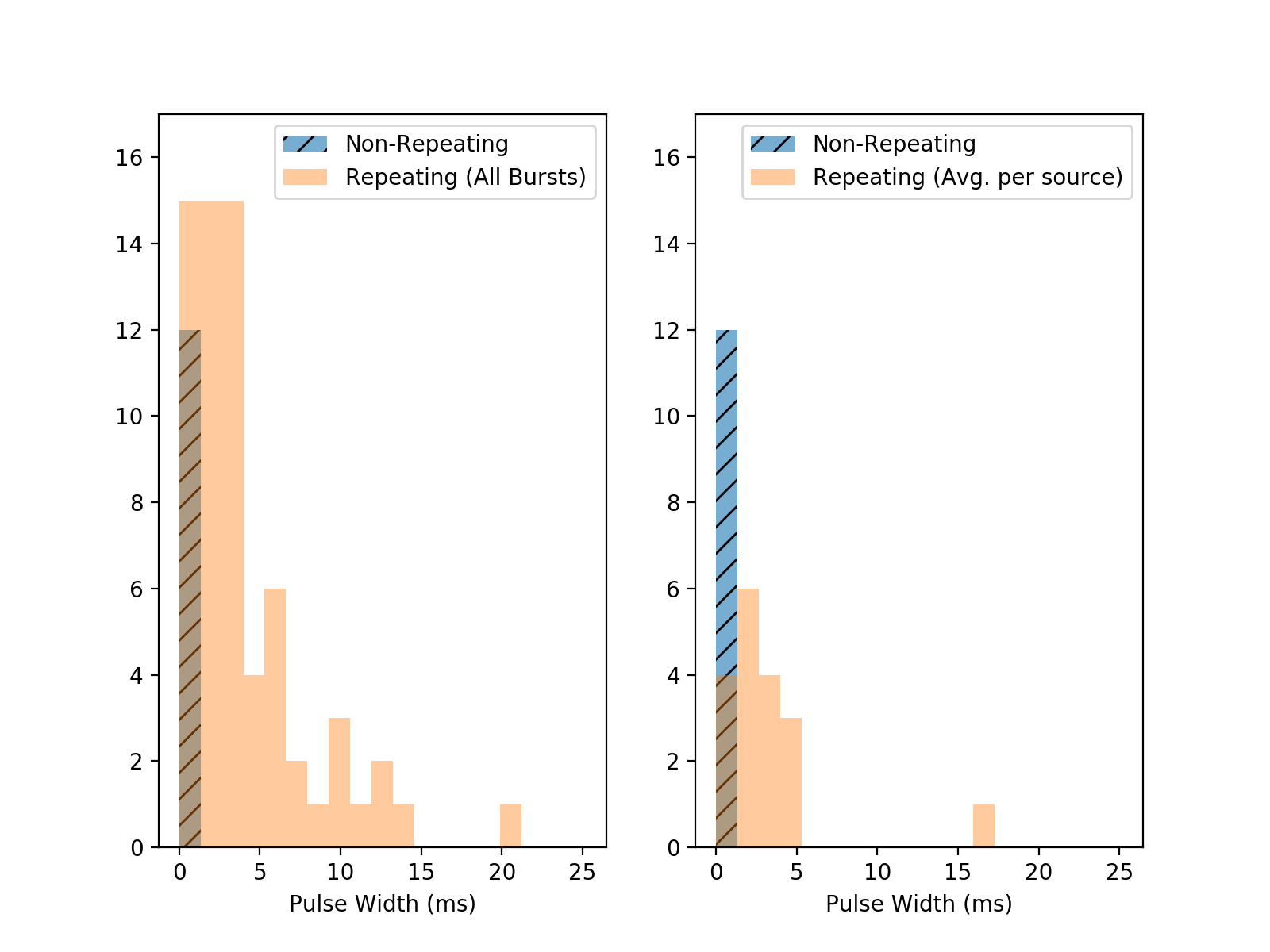}
	\end{center}
\figcaption{Distribution of intrinsic temporal widths for repeating and thus far non-repeating FRB sources observed in the frequency range of 400--800 MHz. For repeating FRBs, the left panel shows the distribution of widths of the Gaussian spectral components for all bursts from each source while the right panel shows only the weighted average of the widths for each source.}
\label{fig:widths}
\end{figure}

Additionally, we included in the comparison only bursts with a detection S/N $>10$, which was the threshold for saving intensity data to disk at the time of detection of the non-repeating sources we considered. This criterion ensures that the two samples have an identical selection function, but omits Source 3 from this analysis since all bursts from that source had S/N $<10$. 
{\ef We compare temporal widths of different Gaussian components of the bursts, including several measurements for any burst with multiple components in the samples used for the comparison. These width distributions are shown in Figure~\ref{fig:widths}.} Using both the Kolmogorov-Smirnov and Anderson-Darling tests, we found that the two samples are not drawn from the same distribution with $\sim5\sigma$ and $\sim4\sigma$ significance, respectively. {\ef When} combining width measurements of different bursts {\ef (or their components)} from each repeating source using inverse-variance weighting, {\ef in order to perform this comparison strictly between FRB sources}, we also found that the difference persists and supports the notion of possibly different emission mechanisms between repeating and thus far non-repeating FRBs. {\ef Alternatively, significant variation in pulse widths could also be caused by differences in the properties of the circumburst media between repeating and thus far non-repeating FRBs. However, this is a less likely possibility considering that the two DM distributions do not have any statistically significant differences.}

As a population, the thus far non-repeaters reported in \citet{abb+19a} show significant scattering, with half having scattering times of $\gapp 1$~ms at 600 MHz. By contrast, only six of the eighteen published CHIME/FRB repeating sources have statistically significant scattering measurements in the 400--800 MHz band whereas the remaining twelve sources have reported upper limits (\citealt{jcf+19}, \citealt{abb+19b}, Paper I). {\ef However, as we described in Section \ref{subsec:burstmorph} and in Paper I, the complex morphology of repeating FRB sources makes their accuracy less transparent. We nonetheless use scattering estimates for CHIME/FRB repeating FRBs at face value in the following statistical analysis.} Here we assumed that the scattering properties of the intervening medium do not change over the period of several months between detection and used the strictest constraint on scattering time listed in Table \ref{ta:bursts} as the measured value for each source. However, several of the reported upper limits are much larger than the measured scattering times for the thus far non-repeaters. This is expected given the above-mentioned width disparity which makes it harder to detect scattering at the $\sim$1-ms level in the repeating bursts.

The lack of measured scattering timescales for several of the repeating sources makes it difficult to directly compare their scattering properties with those of the thus far non-repeaters. Instead, we compared the two distributions with the methodology used in Paper I which sets the reported upper limit on the scattering time for each source to be the $1\sigma$ confidence interval of a normal probability density distribution. A cumulative distribution function for a mixture distribution of the repeater scattering timescales was generated by adding the probability density distributions for all sources. This cumulative distribution function was then compared with measured scattering times for the non-repeaters using a Kolmogorov-Smirnov test. We found that we cannot rule out the possibility of the two samples being drawn from the same parent distribution. We will repeat this analysis once a larger sample of repeater bursts with statistically significant measurements of scattering timescales is obtained.

% the following sentence was originally placed before the final sentence of the above paragraph:
%Since the two samples are statistically indistinguishable, we did not repeat the population synthesis analysis used to interpret the distribution of scattering timescales of the apparent non-repeaters in \citet{abb+19a} with this sample.

Paper I noted a statistically marginal, negative correlation between peak FRB flux and width based on the repeating FRB sample published therein. We repeated this analysis, combining measurements presented in Table \ref{ta:bursts} of the present work to the sample published in Paper I. We found no evidence for correlation between peak fluxes and widths from this enlarged sample. However, as in Paper I, we note that various sources of selection bias have not yet been quantified, and future analyses of such correlations may still be worthwhile.

The 190213 detection of Source 3 at $\sim 1$ ms time resolution consists of two sub-bursts, separated by $\sim 19$ ms without an apparent ``bridge`` in emission in time or frequency -- similar to the 181019 detection of Source 1 in Paper I. Whether a detection like this constitutes one burst comprised of two sub-bursts or two separate bursts remains to be seen; the interpretation is likely model-dependent.

In Table~\ref{ta:bursts}, we reported new measurements of sub-burst linear drift rates. However, in contrast to Paper I, where almost half of bursts showed visible drift, only five of the bursts discussed in this work have significant drift rates. The difference is most likely due to more bursts in the present sample having lower S/N than in Paper I, thus making significant detection of drifting sub-bursts more difficult. The measured linear drift rates appear to be drawn from a similar distribution as in Paper I, and the notion that repeating sources of FRBs exhibit drifting sub-bursts with linear drift rates of order few to tens of MHz/ms in the 400--800 MHz band remains valid. The apparent differences in drift rate uncertainties reported here and in Paper I are due to i.) a non-linear relation between the measured rotation $\theta$ of a two-dimensional Gaussian and the drift rate df/dt $= 1/\tan(-\theta)$, leading to measurements of higher drift rates being more uncertain and ii.) the drift rate uncertainties were determined after marginalising over DM, with the DM uncertainties of the bursts for which we measure drift rate here all being only 0.2--0.3 pc cm$^{-3}$.

As CHIME/FRB total-intensity data have relatively coarse time resolution, it is possible that apparently simple bursts would show substructure and measurable drift if examined at higher time resolution. {\ef This circumstance has been seen in bursts from FRB 121102 \citep{hss+19}}. In addition, the emission mechanism and/or local environments likely introduce burst-to-burst fluctuations to the observed morphology. %Analyses of burst morphology on additional CHIME/FRB bursts that have not repeated thus far, as well as studies of available baseband data, will be the subject of future studies.

As with the sources in Paper I, we found that all bursts in this present sample of repeating sources possess small emission bandwidths of 100--200 MHz{\ef , similar to behavior seen in FRB 121102 \citep{gms+19}}, whereas a large fraction of non-repeating FRBs span the full CHIME band. Given the large sample of repeat bursts obtained with CHIME/FRB, detected over a wide range of beam positions, it is likely that these small emission bandwidths reflect real, intrinsic differences in emission mechanisms and/or local environments between repeating and non-repeating FRB sources. However, a robust statistical analysis of spectral features requires a sample of thus far non-repeating FRBs larger than that considered in Paper I. %We will revisit this question once additional analyses of non-repeating FRBs have been completed.

\subsection{Implications of RM Values}
\label{subsec:RMimplications}

Using CHIME/FRB baseband data, we measured RM values from two new repeating sources. These results are summarised in Table~\ref{tab:RMs}, along with with a previous RM detection from a different CHIME/FRB repeater (Paper I; see Table~\ref{tab:RMs}). RM values reported are those derived from the Stokes QU-fitting method, which were found to give marginally higher linear polarization fractions after de-rotation. {\ef We did not apply any redshift corrections to the values presented in Table \ref{tab:RMs} as their redshifts are not currently known, and the RM contributions from local environments are not constrained.}

We also computed the RM {\ef contributions} from the Milky Way foreground ($\mathrm{RM_{MW}}$), which are provided in Table~\ref{tab:RMs}. We used two methods to determine the Galactic contribution to RM: i) the constructed RM foreground map of \citet{ojg+15}, previously applied in Paper I; and ii) the first Faraday moment of diffuse, polarized emission obtained from the Global Magneto-Ionic Medium Survey (GMIMS) \citep{dlt+19}. The first moment is the mean of the Faraday spectrum, weighted by polarized intensity. Along most high-latitude sightlines the Faraday spectrum is simple, and the first moment is equivalent to the peak Faraday depth. Departures from this scenario arise when polarized emission is Faraday rotated by different amounts. Such departures are commonly observed in diffuse, polarized emission from the Galactic foreground, where its extended nature leads to variable levels of Faraday rotation as a function of {\ef line-of-site} distance. Cases such as this display complex structure of the polarized emission in Faraday depth and can even appear as multiple peaks, preventing an accurate $\mathrm{RM_{MW}}$ value from being inferred. This does not appear to be a problem here, with all repeater directions yielding relatively simple Faraday spectra{\ef , as is also the case for FRB 121102 \citep{msh+18}}.

Significant differences can be seen between certain $\mathrm{RM_{MW}}$ values when using the two methods. Notably, FRB180916's position near the Galactic midplane likely complicates the interpretation of the first Faraday moment from GMIMS. This discrepancy occurs because sightlines at lower latitudes probe a greater extent of the Galactic foreground emission. In low-latitude cases such as this, the first Faraday moment of the emission is unlikely to be an accurate proxy for $\mathrm{RM_{MW}}$. In fact, even idealized scenarios with minimal path length through the foreground and no magnetic field reversals can still produce systematic bias between the first Faraday moment of diffuse, polarized emission and the true $\mathrm{RM_{MW}}$ value \citep{obv+19}. A larger sample of extragalactic polarized sources, including FRBs, will be helpful in identifying well behaved regions of the sky where diffuse, polarized emission can be leveraged to better subtract the foreground RM contribution.

Regardless of the method used for subtracting the RM contribution of the Galactic foreground, a considerable gap ($\mathrm{>10^5 \textrm{ rad }m^{-2}}$) exists between RM values obtained for repeating sources observed by CHIME/FRBwith those associated with bursts from FRB121102. { \ef It is possible that depolarization arising from intrachannel Faraday rotation greatly reduces sensitivity to polarized signal with extreme RMs, effectively preventing detection. At the native channelization of CHIME baseband data, sensitivity to polarized signal drops significantly for RMs beyond $\sim10^3 \textrm{ rad }m^{-2}$, with the exact value depending on the specifics of the burst spectrum. However, even after accounting for this possibility, repeater RMs discovered with CHIME appear rather modest compared to the effective range over which RM detections are possible. At the very least, these results demonstrate that RM is not a perfect discriminant between repeating and (apparently) non-repeating sources and further highlight the distinct nature of FRB 121102. 

In addition to RM, polarization fraction (both linear and circular) and the polarization position angle (PA) across the burst phase offer additional diagnostic information for informing FRB emission models (see Paper I for details). The flat PA curves for Sources 2 and 6 shown in Figure \ref{fig:pol_profs} are consistent with those of FRB 180916.J0158+65 (Paper I) and FRB 121102 \citep{ssh+16a,msh+18,gsp+18,hss+19}; yet this does not appear to be a common feature of all FRB sources with FRB 110523 displaying evidence for PA variation across the burst duration \citep{mls+15a}. The significance of this burst being from a thus far non-repeating source remains to be seen and encourages future polarized FRB observations. 

The nearly $\sim100\%$ linear fraction polarization of Source 2 and the much lower $>20\%$ of Source 6 are largely consistent with the level of heterogeneity expected from the published sample. However, we urge caution when interpreting the fractional polarizations of Source 6 as significant instrumental effects lead to substantial mixing of the Stokes parameters, yielding misleading linear and circular polarization fractions. However, these polarized instrumental effects are highly dependent on position in the primary beam of the telescope and for Source 2 appear to be sub-dominant. Nonetheless, analysis techniques for correcting the differential polarized response across the primary beam are currently in development and will be reported elsewhere.} 

{ \ef The combined polarization information of these new repeating FRB sources will likely motivate revision} of existing FRB emission models, many of which use polarization properties of FRB 121102 as key information in their initial construction. An example of such a case is the young magnetar model of \citet{mm18}, which posits a dense, ionized nebula around a central engine giving rise to the high RM observed in FRB 121102. A variant of this model \citep{mms19} has since been developed and found to be consistent with burst properties reported for the initial batch of CHIME repeaters (see Paper I). Moreover, this model allows for a range of RM values, dependent on the age of the magnetar and the specifics of its formation channel \citep{mbm19}. According to this scenario, the RM should decay monotonically with time \citep{pg18} giving rise to a trend of younger, more active sources with higher RMs and older, less active sources with lower RMs. The fact that CHIME seems to preferentially observe repeaters with relatively low RM values is at odds with this prediction but can possibly be explained by invoking a concomitant evolution of the peak flux frequency as a function of age, such that CHIME/FRB tends to detect older repeating sources \citep{klb17, mms19}. The detection of only one burst from FRB121102 in the CHIME band supports this assertion, but a much larger sample of repeater RMs over a wide range of bandwidths will be needed to make any robust claims.

Alternatively, it is also possible that FRB 121102's large RM is not from an associated nebula but from highly magnetized environment commonly found in dense, star-forming regions and/or around massive accreting black holes {\ef \citep[see discussion in][]{msh+18}}, as in the case of the Galactic-center magnetar \citep{dep+18}. In this framework, the circumstance of FRB 121102 residing in such environments can be due purely to chance, or is perhaps a reflection of the precursor's preference for these environments. In either case, the observed RM is independent of the central FRB engine driving emission and therefore not directly correlated with properties relating to its emission. This interpretation can be tested in the future with a much larger RM sample combined with interferometric localization and multi-wavelength observations capable of probing the local environment. 

In addition, multi-epoch observations of newly discovered repeaters may provide evidence for secular evolution of the RM similar to that seen for FRB 121102 \citep{msh+18,gsp+18}. The existence of such a trend would be difficult to reconcile in this framework and would more strongly support emission models that explicitly link evolution of local environment with the FRB engine itself.

\subsection{Repetition Rates}
\begin{figure}
\includegraphics[width=0.95\textwidth]{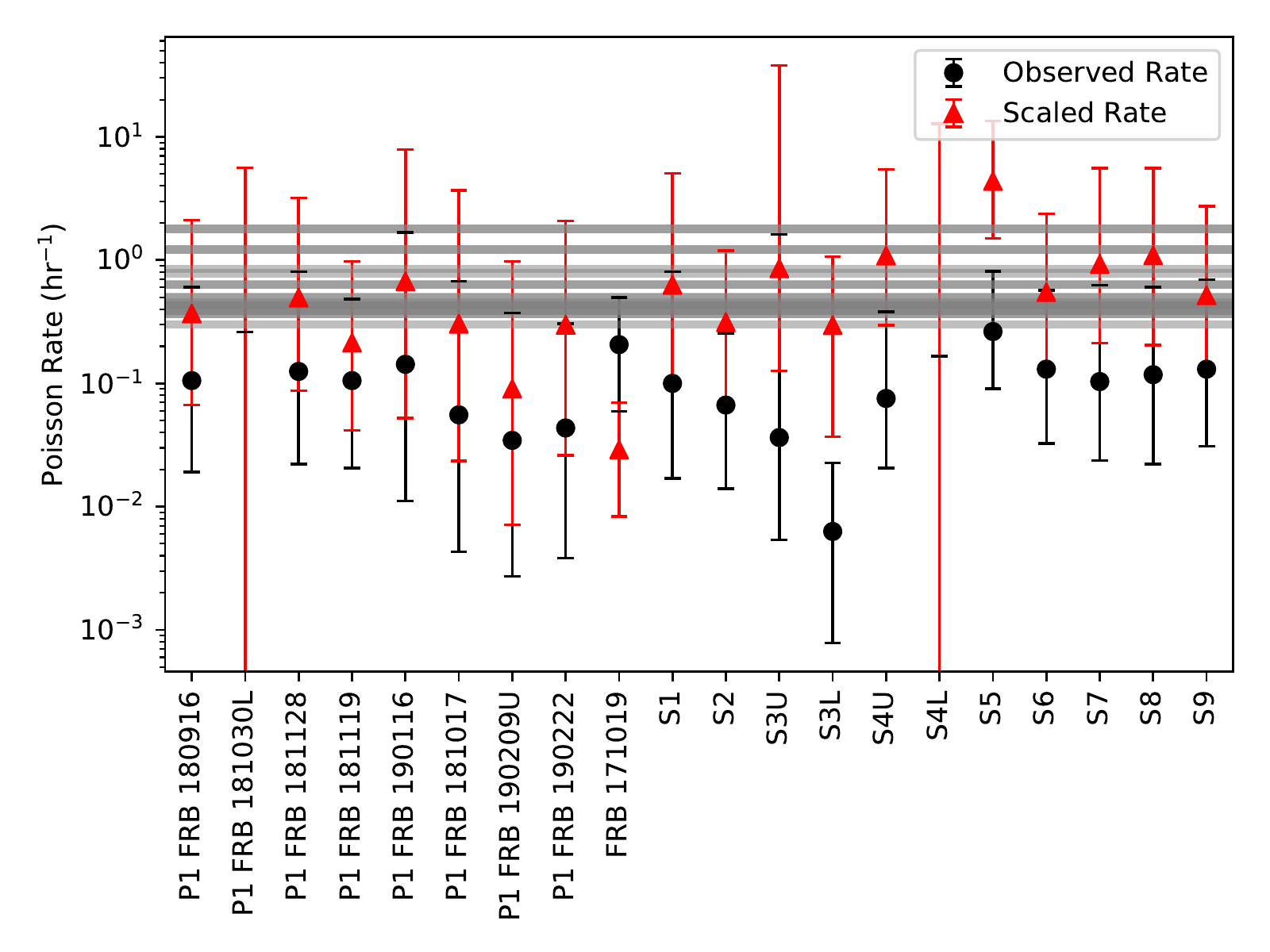}
\caption{Repetition rates of repeaters and upper limits for single bursts from ASKAP \citep[gray lines; ][]{smb+18}. Observed rates are denoted as black circles, while rates scaled by fluence sensitivity are shown as red triangles. Repeaters from Paper I are labelled ``P1". Repeaters from Table~\ref{ta:repeaters} are labelled as ``S\texttt{x}". The repetition rate for FRB 171019 is calculated from the GBT 820 MHz observations \citep{kso+19}. The rates (red triangles) and upper limits are scaled to a fluence limit of 1 Jy--ms using a scaling of $S^{1.5}$ where $S$ is the sensitivity of the search. The ``U" and ``L" suffixes for CHIME/FRB-detected repeaters denote the rates calculated from number of detections, exposure and sensitivity in the upper and lower transits, respectively, for circumpolar sources. 1-$\sigma$ error bars are shown assuming a Poisson distribution.} 
\label{fig:repetition_rates}
\end{figure}

We calculated the repetition rates of these repeaters and compared them to the limits on single bursts observed by {\ef the Australian Square Kilometre Array Pathfinder} \citep[ASKAP;][]{smb+18}. {\ef Note that a repeat burst was found from FRB 171019, one of the sources from \citet{smb+18}, in follow up observations at 820 MHz with the Green Bank Telescope \citep{kso+19}.} In order to compare the repetition rates across different fluence limits, we scale the observed rates (based on the detections and exposure in Table~\ref{ta:repeaters}) by $S^{1.5}$ where $S$ is the fluence sensitivity (also specified in Table~\ref{ta:repeaters}) and the exponent of 1.5 reflects an assumed Euclidean spatial distrbution of sources. We calculated the rates for upper and lower transits separately for sources that are circumpolar in the CHIME/FRB field of view.  Figure~\ref{fig:repetition_rates} shows the observed and scaled rates of repetition for repeaters from Paper I, this work, {\ef and from \citet{kso+19}} as well as scaled upper limits on repetition rates for the single bursts reported by \citet{smb+18}. We found that most of the repetition rates for repeaters are at or lower than the 1-$\sigma$ upper limits from the ASKAP observations. The scaled repetition rate of Source 5 is marginally higher than most of the upper limits from ASKAP observations.

\subsection{Multi-wavelength Follow-up}

We checked for any catalogued ionized regions \citep{anderson2014wise,green2019revised} or star-forming regions \citep{avedisova2002catalog,rice2016uniform} within the Milky Way galaxy that are coincident with the localization areas of the 9 new repeaters presented in Table \ref{ta:repeaters}, and found none. We also estimated the maximum redshift ($z_{\rm max}$) limit for all FRBs to identify plausible galaxy candidates. To estimate the Milky Way (MW) contribution to the observed DMs, we considered the smaller of NE2001 \citep{ne2001} and YMW16 models' \citep{ymw17} predicted Galactic DM values, and added a MW halo contribution of 50 pc cm$^{-3}$ \citep{prochaska2019probing}. We then subtracted this value from the observed DMs to estimate the extragalactic DM contribution. These excess DMs were converted to z$_{\rm max}$ by using the DM-redshift relation: DM$_{\text{\ef ex}}$ $\approx 900z$ \citep{zhang2018fast}. These redshift estimates are approximate upper limits as we did not account for the DM of the host galaxy. Using these $z_{\rm max}$ values, we searched various catalogues of nearby galaxy clusters \citep{abell1989catalog,bohringer2000northern,wen2017catalogue,hao2010gmbcg} and found none within the localization regions of the FRBs.
%\vspace{8pt}

As our repeating FRBs have poor localization and large estimated $z_{\rm max}$, the chance coincidence probability of finding even a massive star-forming galaxy is large. This circumstance remains true for Source 4 despite possessing the lowest DM excess among our repeating FRB sample, DM$_{\rm ex} \approx {\ef 115} \textrm{ pc cm}^{-3}$. However, the largest RM excess in our sample, $|{\rm RM}_{\rm ex}|$ $\approx$ 490 rad m$^{-2}$ for Source 6, likely suggests {\ef considerable} host DM contribution and, hence, a nearby host galaxy. This fact prompted us to look for plausible host galaxy candidates around Source 6, which has the second lowest extragalactic DM among Sources 1-9 (DM$_{\rm ex}\approx$ 140 pc cm$^{-3}$). We estimated z$_{\rm max} \approx$ 0.16 for Source 6 and found a pair of face-on star-forming merging galaxies, SDSS J135159.17+480729.0 and SDSS J135159.87+480714.2, at spectroscopic redshift = 0.064 \citep{alam2015eleventh}; these galaxies possess the lowest $z_{\rm max}$ among the catalogued galaxies within the localization region of Source 6, and either one can easily account for the observed excess DM. Using the luminosity function of massive galaxies from \cite{faber2007galaxy}, we estimate the density of galaxies with M$_{B} < -$20 to be $\sim$ 0.0015 Mpc$^{-3}$. By assuming a pair merger fraction = 0.01 \citep{bell2006merger}, we estimated the probability of finding a pair of massive merging galaxies by chance to be  $< 5\%$ within the Source 6 containment region for $z_{\rm max} =$ 0.16. Therefore, we consider these galaxies to be interesting candidates for the host of Source 6.

\section{Conclusions}
\label{sec:conclusions}

We have reported on the discovery of nine new repeating FRB sources from CHIME/FRB. Multiple bursts from these nine sources were collected during a $\sim$1 yr period of telescope operation, and display complex spectro-temporal behavior similar to previously reported repeating FRB sources {\ef \citep[e.g., Paper I,][]{hss+19}}. We found that the statistical properties of the data set presented in this work confirm the findings of Paper I: the DM distributions of repeating and thus far non-repeating CHIME/FRB sources are indistinguishable, while the distributions of temporal widths between the same populations are statistically different at the $4\sigma$ level, with the repeating-source population producing larger widths. As first discussed in Paper I, this observation likely indicates intrinsic differences between emission mechanisms and/or environments local to the sources that produce repeating and thus far non-repeating bursts. {\ef Future studies of these properties from the $\sim$700 bursts observed by CHIME/FRB will place stronger constraints on these statistical differences and are forthcoming.}

Analysis of baseband data acquired for two of these sources yielded significant estimates of Faraday rotation, with RM = $-$20(1) rad m$^{-2}$ for Source 2 and RM = $-$499.8(7) rad m$^{-2}$ for Source 6. {\ef The large RM difference between bursts associated with FRB 121102 and those reported here further indicates a uniqueness in the local magnetized environment of FRB 121102. Furthermore, the relatively modest RMs discovered with CHIME/FRB do not yet indicate clear  differences with the non-repeating sample. This suggests that RM is perhaps a poor discriminant for repeating and thus far non-repeating FRB sources, but any robust statistical claim requires a larger sample.} When comparing to expected Galactic contributions, the low excess DM and large excess RM for Source 6 are suggestive a potentially nearby host galaxy with possible star formation; we searched for candidates and identified a pair of merging galaxies at redshift $z = 0.064$ as a tentative candidate host galaxy. A more precise localization of Source 6 is needed to confirm or exclude this tentative hypothesis.

To date, CHIME/FRB has discovered a total of 18 repeating FRB sources \citep[see][Paper I and this work]{abb+19b} {\ef out the 20 currently known, with two other repeating sources found using the Arecibo Observatory \citep{ssh+16a} and ASKAP \citep{kso+19}}. Each source presents an opportunity for arcsecond localization with large telescope arrays and, once achieved, multi-wavelength studies of the host galaxy and any associated intra-galactic environment. {\ef Indeed, i}nterferometric follow-up of several CHIME/FRB repeating sources are underway with the {\tt realfast} backend at the VLA \citep{lbb+18} and the European VLBI Network \citep[EVN;][]{mph+17}{\ef ; the EVN recently localized FRB 180916.J0158+65, discovered by CHIME/FRB (Paper I), with mas precision to a star-forming region in a nearby massive spiral galaxy \citep{mnh+20}.} We encourage additional community involvement, both in localizing these sources and targeting localized sources for constraining emission and activity across the {\ef electromagnetic} spectrum.

\bigskip

\acknowledgements
We thank the Dominion Radio Astrophysical Observatory, operated by the National Research Council Canada, for gracious hospitality and useful expertise. The CHIME/FRB Project is funded by a grant from the Canada Foundation for Innovation 2015 Innovation Fund (Project 33213), as well as by the Provinces of British Columbia and Qu\'ebec, and by the Dunlap Institute for Astronomy and Astrophysics at the University of Toronto. Additional support was provided by the Canadian Institute for Advanced Research (CIFAR), McGill University and the McGill Space Institute via the Trottier Family Foundation, and the University of British Columbia. The Dunlap Institute is funded by an endowment established by the David Dunlap family and the University of Toronto. Research at Perimeter Institute is supported by the Government of Canada through Industry Canada and by the Province of Ontario through the Ministry of Research \& Innovation. The National Radio Astronomy Observatory is a facility of the National Science Foundation (NSF) operated under cooperative agreement by Associated Universities, Inc.

M.~B. is supported by a Doctoral Research Award from the Fonds de Recherche du Qu\'ebec: Nature et technologies (FQRNT). P.~C. is supported by an FRQNT Doctoral Research Award. V.~M.~K. holds the Lorne Trottier Chair in Astrophysics and Cosmology and a Canada Research Chair, receives support from a Discovery Grant of the Natural Sciences and Engineering Research Council of Canada (NSERC) and Herzberg Award, as well as from an R. Howard Webster Foundation Fellowship from CIFAR, and from the FRQNT Centre de Recherche en Astrophysique du Qu\'ebec. D.~M. is a Banting fellow. Z.~P. is supported by a Schulich Graduate Fellowship. M.~D. is supported by a Killam Fellowship and receives support from an NSERC Discovery Grant, CIFAR, and from the FRQNT Centre de Recherche en Astrophysique du Qu\'ebec. B.~M.~G. acknowledges the support of NSERC through grant RGPIN-2015-05948, and of the Canada Research Chairs program. SMR is a CIFAR Fellow and is supported by the NSF Physics Frontiers Center award 1430284. P.~S. is a Dunlap Fellow and an NSERC Postdoctoral Fellow. FRB work at UBC is supported by an NSERC Discovery Grant and by CIFAR. The baseband voltage system is funded in part by a John R. Evans Leaders Fund award, from the Canadian Foundation for Innovation, to I.H.S.

\bibliographystyle{aasjournal}

\bibliography{main.bbl}
%\bibliography{frbrefs,modrefs,psrrefs,crossrefs,journals1}

%\end{thebibliography}

\appendix
\section{Chance Coincidence Probabilities with large numbers of FRBs}
\label{sec:chance}
Once a large sample of FRBs has been detected, the probability of identifying two bursts with similar DM and sky location can become non-negligible. In such circumstances there are two questions that must be answered: 

\begin{enumerate}
\item Given $N$ FRBs detected by a survey, what is the probability that \emph{any} two FRBs will lie within the same phase-space bin of dimension ($\Delta\mathrm{DM},\,\Delta\mathrm{R.A.}\cos(\mathrm{Dec.}),\,\Delta\mathrm{Dec.}$)? We refer to this as the ``global probability."
\item If we detect two FRBs in the same ($\Delta\mathrm{DM},\,\Delta\mathrm{R.A.}\cos(\mathrm{Dec.}),\,\Delta\mathrm{Dec.}$) bin during our survey, what is the probability that they are physically unrelated? We refer to this as the ``individual probability."
\end{enumerate}

Here we answer both questions for all CHIME/FRB repeaters found to date (i.e., from this work and from Paper I), given that CHIME/FRB has detected $\sim700$ FRBs during the observing period mentioned in Section \ref{sec:obs}. A detailed analysis of the $\sim700$ FRBs is underway and will be published elsewhere. Here we limit our discussion to the declinations and DM distribution of the FRBs in the CHIME/FRB survey; we marginalised over the R.A. dimension for reasons discussed below. 

\subsection{Global probability}
The first question is mathematically equivalent to the non-uniform birthday problem{\ef, i.e.,} what is the chance that two people in a group of $N$ share the same birthday. The uniform case, in which each birthday (or each $\Delta\mathrm{DM},\,\Delta\mathrm{R.A.}\cos(\mathrm{Dec.}),\,\Delta\mathrm{Dec.}$ bin) are equally likely, is straightforward to solve. The solution for the non-uniform case has been described in terms of recursion relations for a coincidence of two among $N$ objects \citep{mase1992}, and for a general case of $m$ among $N$ objects \citep{sandell1991}. However, for the case of CHIME/FRB, the parameter space is large enough that the recursion formulae and their approximations are not computationally tractable. Instead, we simulated the current results of the CHIME/FRB survey using Monte Carlo methods in order to understand how often we misidentify repeaters with a given criteria. 

Following the method in Appendix A of Paper I we estimated the detection probability distribution function from the $\sim$700 FRBs found by the CHIME/FRB survey. This calculation incorporated variations in exposure times, sensitivity at different declinations and DMs. We assumed that the probability of detection is independent of R.A. and local time since the survey has been operating for over a year and any variation in R.A. would have been averaged out. The probability distribution was smoothed by a Gaussian kernel with a 350 $\mathrm{pc\,cm^{-3}}$ DM scale and 10\degrees\ angular scale in declination.  

%\textbf{Show histogram of trials.}
\begin{figure}
    \includegraphics[width=0.95\textwidth]{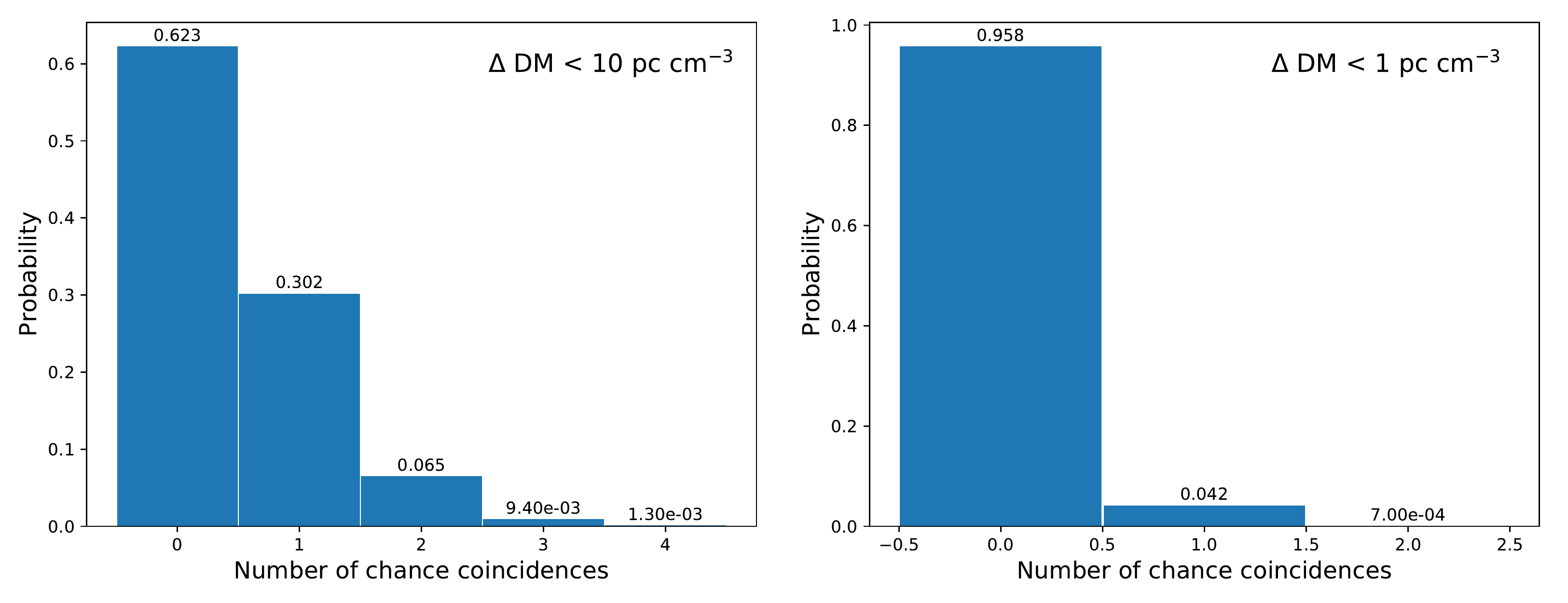}
    \caption{Histogram of the MC trials as a function of the number of detected chance coincidences. Both trials had the same localization criteria, $\Delta\mathrm{R.A.}\cos(\mathrm{Dec.})<1\deg ,\,\Delta\mathrm{Dec.}<1\deg$. The left panel displays a DM {\ef criterion} of $\Delta\mathrm{DM} < 10\,\mathrm{pc\,cm^{-3}}$, and the right panel displays a DM {\ef criterion} of $\Delta\mathrm{DM} < 1\,\mathrm{pc\,cm^{-3}}$. A stricter DM cutoff allows for fewer chance coincidence detections when sampling 700 independent FRBs.}
\label{fig:mc_distributions}
\end{figure}

We sampled 700 independent FRBs from this probability distribution for 10$^4$ trials and determined how many FRBs would be counted as repeaters with the following criteria: I) $\Delta\mathrm{DM} < 10\,\mathrm{pc\,cm^{-3}},\,\Delta\mathrm{RA}\cos(\mathrm{Dec.})<1\deg ,\,\Delta\mathrm{Dec.}<1\deg$ and II) $\Delta\mathrm{DM} < 1\,\mathrm{pc\,cm^{-3}},\,\Delta\mathrm{R.A.}\cos(\mathrm{Dec.})<1\deg ,\,\Delta\mathrm{Dec.}<1\deg$. Figure~\ref{fig:mc_distributions} displays a histogram of the trials as a function of number of detected chance coincidences for the two criteria. For the first criterion, $\sim$40\% of the trials yield at least one set of independent FRBs falsely identified {\ef as coming from} the same repeating source. With the second {\ef criterion}, the probability is $\sim$4\%.

We then determined the selection criteria to use, such that our repeaters have a chance coincidence probability ($p_{\rm CC}$) of $\sim$1 in 100 simulations. We found: i) given a localization uncertainty of $\sim$20', the DM tolerance has to be $\Delta\mathrm{DM} < 2.0\,\mathrm{pc\,cm^{-3}},$ and ii) given a DM tolerance of $10.0\,\mathrm{pc\,cm^{-3}}$, the spatial localization has to be $\Delta\mathrm{R.A.}\cos(\mathrm{Dec.})<0.1\deg ,\,\Delta\mathrm{Dec.}<0.1\deg$. These criteria were found before looking at the properties of the repeating FRB candidates.

We note, however, that the probability distribution was smoothed in both declination and DM, and assumed to be uniform in R.A. Any clustering or non-uniformity at scales smaller than the smoothing scales would increase the $p_{\rm CC}$. Additionally, the results of our simulations imply that all future repeating CHIME/FRB candidates must have extremely spatially well-localized bursts to keep our chance coincidence probabilities low. A source with minimal spatial uncertainties would also allow for more DM evolution across the bursts{\ef , which has been seen in FRB 121102 at the $\sim$1\% level \citep{hss+19,jcf+19}}.

\subsection{Individual Probabilities}
The above simulations do not identify which bursts might be misidentified as repeaters. If $p$ is the probability of detecting a burst in the ($\Delta\mathrm{DM},\,\Delta\mathrm{R.A.}\cos(\mathrm{Dec.}),\,\Delta\mathrm{Dec.}$) neighborhood of a known burst, the probability that one of the $n$ subsequent bursts will be in the same bin by coincidence is $1-(1-p)^n$. Since these repeaters are identified from a population of $N\approx700$ FRBs, we use a trials factor of 700 to calculate the final coincidence probabilities.  

Here, we considered only the repeaters with two detected bursts, since the chance coincidence probabilities for three or more bursts occurring within the same phase-space bin are negligible for our current sample size. We started with the probability distribution above, normalised over the full sky and DM range. Table~\ref{tab:rep_pcc} specifies the candidate repeaters with two bursts and the range of $\Delta\mathrm{DM},\,\Delta\mathrm{R.A.}\cos(\mathrm{Dec.}),\,\mathrm{and}\,\Delta\mathrm{Dec.}$ for the detected bursts.  For Sources 1 and 2, the trials-adjusted chance coincidence probability ($Np_{\rm CC}$) remains below the 0.1\% level, indicating robust association of repeat bursts despite the large CHIME/FRB sample. However, $Np_{\rm CC}$ = 10\% for Source 8, which makes this association \st{statistically less significant. Future detections of bursts from Source 8 will confirm its status as a repeater.}
%the most marginally significant of the present work.}

\begin{table}[t]
\begin{center}
\caption{Chance Coincidence Probabilities for Repeating FRB Candidates with Two Bursts}
%\footnotesize
\centering
 
%\hspace{-1.8in}
\begin{tabular}{cccccccc} \hline
    Source & Name$^a$  &  $\Delta\mathrm{R.A.}\cos(\mathrm{Dec.})^b$ & $\Delta$Dec.$^c$ & $\Delta$DM$^d$ & $n_\mathrm{int}^e$ &  $p_\mathrm{CC}$ & $N_\mathrm{trials}p_\mathrm{CC}$ \\
          &       &  (deg)    &  (deg) & (pc~cm$^{-3}$) &   &    &      \\\hline
1 & 190208.J1855+46 & 0.5 & 0.5 & 1 & 101 & $7.4\times10^{-6}$ & $5\times10^{-3}$ \\
2 & 190604.J1435+53 & 0.5 & 0.5 & 1 & 5 &  $3.7\times10^{-7}$ & $3\times10^{-4}$ \\
8 & 190212.J02+20   & 2.2 & 0.5 & 1 & 137 & $1.5\times10^{-4}$ & 0.1\\\hline
%9 & 190303.J1221+70 & 1.6 & 0.5 & 4 & 201 & $6.3\times10^{-4}$ & 0.4\\\hline
\end{tabular}

\label{tab:rep_pcc}
\end{center}
$^a$ As defined in Table~\ref{ta:repeaters}.\\
$^b$ Maximum of beam radius (0.5$^\circ$) or R.A. spread of the bursts (as denoted in Figure~\ref{fig:localization}). \\
$^c$ Maximum of beam radius (0.5$^\circ$) or Dec. spread of the bursts. The beam area is calculated as $\pi\Delta\mathrm{R.A.}\cos(\mathrm{Dec.})\times\Delta\mathrm{Dec}$.\\
$^d$ Maximum of 1 pc~cm$^{-3}$  or DM spread of the bursts. \\
$^e$ Number of FRBs detected in the interval between the two bursts. \\
\end{table}
\normalsize

\section{Exposure Estimation}
\label{app:exposure}
\begin{figure}
\includegraphics[width=0.95\textwidth]{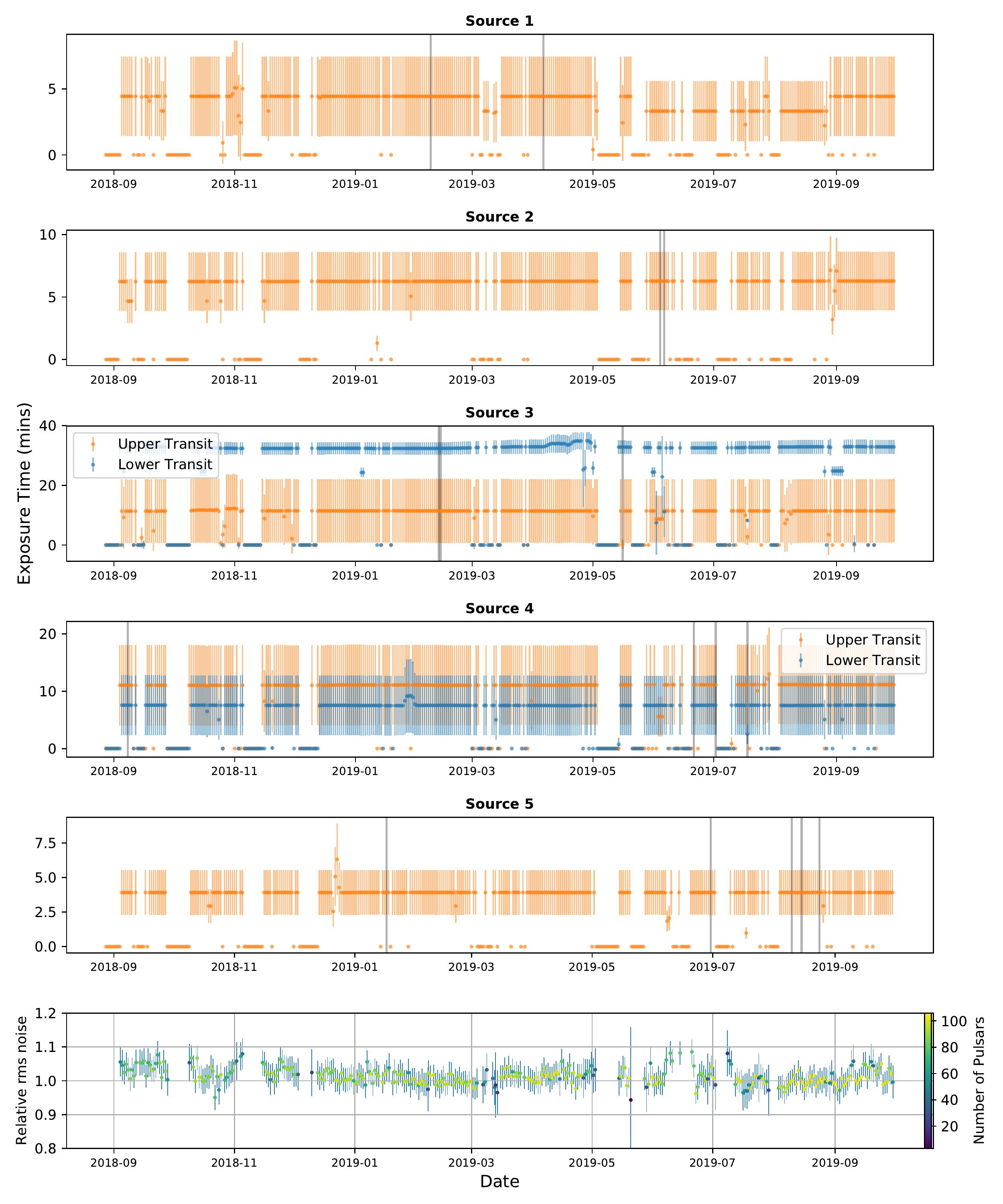}
\caption{Timeline of CHIME/FRB's daily exposure to the new repeating FRB sources for upper and lower transits, if observable. Days on which a burst was detected are indicated by solid lines. The errors on the exposure are due to uncertainties in the source positions. The increase in exposure time from its typical value for some of the days is due to the occurrence of two transits in the same solar day caused by the length of a solar and a sidereal day being slightly different. The RMS noise is estimated using pulsars detected by CHIME/FRB for each sidereal day for which the telescope was operating with the same gain calibration. The RMS noise for each pulsar is measured relative to the median over all days the pulsar was detected with the weighted average of measurements from several pulsars (the number of which is denoted by the marker colors) plotted here.}
\label{fig:exposure}
\end{figure}

The timeline of the exposure of the CHIME/FRB system to each of the sources presented in this work is plotted in Figure \ref{fig:exposure}. The exposure is calculated for the transit of each source across the FWHM region of the synthesized beams at 600 MHz and includes transits in the interval from 2018 August 28 to 2019 September 30. We excised transits from the reported exposure for which the RMS noise (shown in Figure \ref{fig:exposure}) was different by more than one standard deviation from the mean RMS noise in the above-mentioned interval. The fraction of excised transits averaged about 5\% for each source. 

The reported exposures only include intervals in each transit for which the detection pipeline was fully operational. The reduction in daily exposure for Sources 1 and 6 for a period of several months can thus be attributed to the failure of the computing node designated to process data for one of the four synthesised beams through which these sources transit. Sources 3 and 4 have declinations greater than +70$^\circ$ thereby allowing both upper and lower transits to be observable with the CHIME/FRB system. However, there were no bursts detected during the lower transit of either of these sources. {\ef Source 3 is particularly interesting since a significant fraction of its 90\% confidence localization region is located between the FWHM regions of two synthesized beams during the upper transit (see Figure 1). Therefore, for this source, all allowed sky locations which transit between the two beams have zero exposure. This circumstance results in an average exposure for Source 3 over the entire positional uncertainty region of $55 \pm 52$ hours, despite its high declination.}

\addtocounter{figure}{-1}
\begin{figure}
\includegraphics[width=0.95\textwidth]{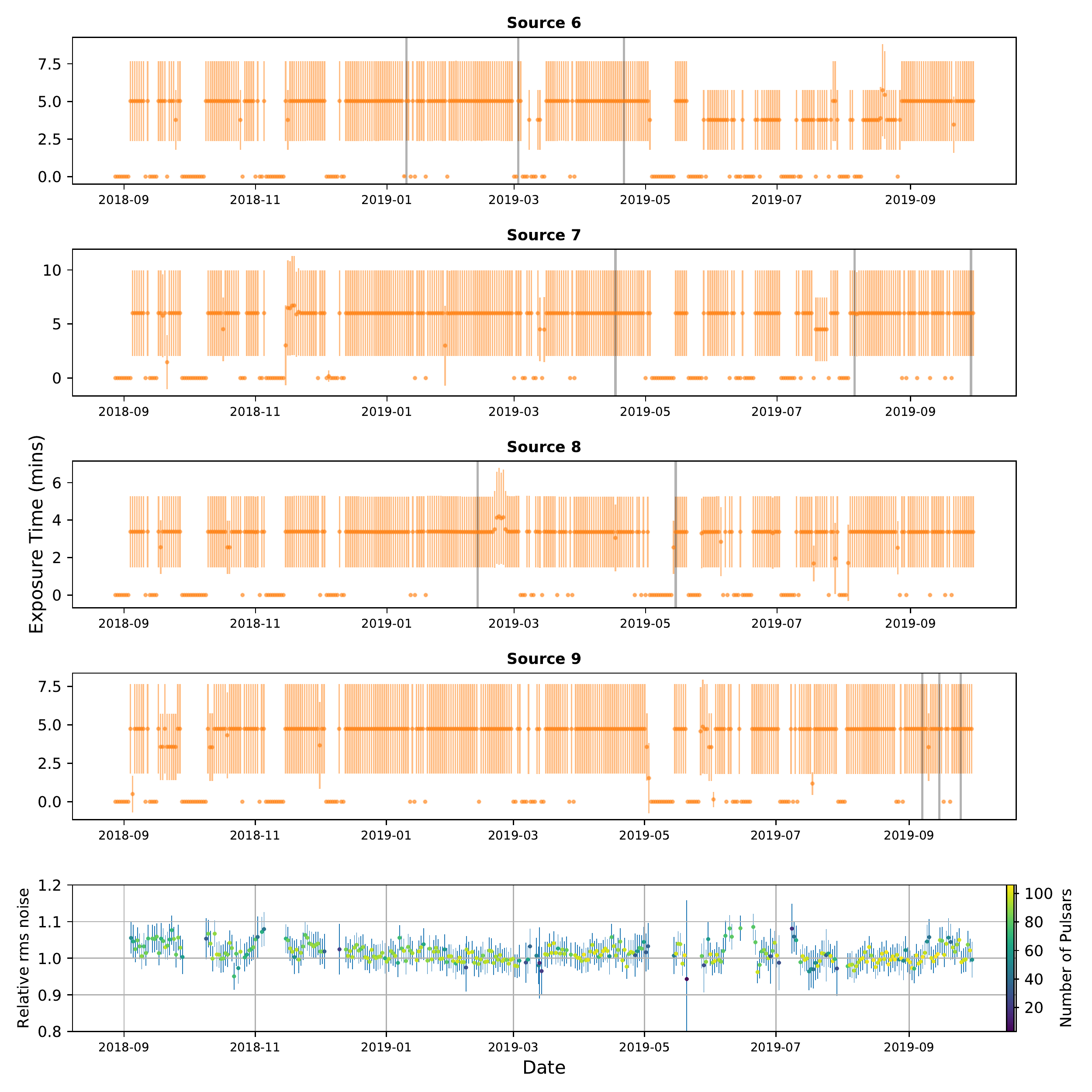}
\caption{Timeline of CHIME/FRB's daily exposure to the new repeating FRB sources. (cont.)}
\end{figure}

\end{document}